\documentclass[final]{jfp-epi}

\usepackage{wrapfig}
\usepackage{subcaption}
\usepackage{tikz}
\usepackage{graphicx}
\usepackage{amsmath}
\usepackage{float}

\usepackage{natbib}
\newcommand{\ttt}[1]{\texttt{\small #1}}

\jfpVolume{36}
\jfpArticle{4}
\jfpDOI{10.46298/jfp.17766}
\jfpYear{2026}
\received[Submitted]{August 2024}
\received[accepted]{December 2025}

\begin{document}




\title{Mixing visual and textual code}


       \author{Leif Andersen}
       \orcid{orcid}
   \affiliation{\institution{University of Massachusetts Boston}
   \city{Boston}
   \state{Massachusetts}
   \country{USA}\\
     \authoremail{leif@leifandersen.net}}

      \author{Michael Ballantyne}
      \orcid{orcid}
  \affiliation{\institution{Northeastern University}
  \city{Boston}
  \state{Massachusetts}
  \country{USA}\\
    \authoremail{ballantyne.m@northeastern.edu}}

      \author{Cameron Moy}
      \orcid{orcid}
  \affiliation{\institution{Northeastern University}
  \city{Boston}
  \state{Massachusetts}
  \country{USA}\\
    \authoremail{camoy@ccs.neu.edu}}

      \author{Matthias Felleisen}
      \orcid{orcid}
  \affiliation{\institution{Northeastern University}
  \city{Boston}
  \state{Massachusetts}
  \country{USA}\\
    \authoremail{matthias@ccs.neu.edu}}

      \author{Stephen Chang}
      \orcid{orcid}
  \affiliation{\institution{University of Massachusetts Boston}
  \city{Boston}
  \state{Massachusetts}
  \country{USA}\\
    \authoremail{stephen.chang@umb.edu}}

\begin{abstract}

The dominant programming languages support nothing but linear text
to express domain-specific geometric ideas. What is needed are
hybrid languages that allow developers to create visual syntactic
constructs so that they can express their ideas with a mix of
textual and visual syntax tailored to an application domain. This
mix must put the two kinds of syntax on equal footing and, just as
importantly, the extended language must not disrupt a programmer's
typical workflow. This means that any new visual syntax should be a
proper language extension that is composable with other language
features. Furthermore, the extensions should also preserve static
reasoning about the program. This paper presents \textsc{Hybrid Clojure\-Script},
the first such hybrid programming language.  \textsc{Hybrid Clojure\-Script} allows
programmers to add visual interactive syntax and to embed instances
of this syntax within a program's text. An enhanced hybrid IDE can
then display these embedded instances as mini-GUIs that
programmers interact with, while other IDEs will show a textual
representation of the syntax.  The paper argues the necessity of
such an extensibility mechanism, demonstrates the adoptability of
the design, and discusses what might be needed to use the design
in other languages.
\end{abstract}

\maketitle

\section{Textual code is insufficient, visual code might be too}\label{sec:intro}

Programming languages help programmers communicate their ideas, both to
computers and to other programmers. Linear text suffices for this
purpose most of the time, but many ideas are inherently geometric and
better expressed visually.

\subsection{Background and research landscape}

Visual programming languages~\cite{bd:visual} offer a potential
solution, but many require all parts of a program to be visual, even
though not all code is best understood this way. Further, as decades of
programming history shows, developers will never completely give up
textual languages---so any graphical syntax must \textit{supplement},
not displace, textual syntax.  Another issue is that many visual
languages are not extensible or
"rich"~\cite{horowitz2023liverichcomposable}; that is, programmers
cannot tailor the graphical syntax to a domain.

The extensive work on live
programming~\cite{sandblocks, horowitz2023liverichcomposable,
livesurvey2018, vivalive1990, bretvictor} has done much to advance
code beyond static text, into the visual world. These systems
typically allow code to be edited visually with feedback from runtime
behavior of the program (thus moving away from both the "static" and
"text" part of "static text"). While this enables a multitude of new
applications, it also comes with the well-understood tradeoffs---e.g.,
in security or static reasoning ability---of overly dynamic code (see
the extensive warnings that have been written about using \ttt{eval}
in JavaScript~\cite{jsgoodparts, evalthatmendo}).

This paper, a substantial revision and extension of the authors' prior
work~\cite{abf:adding}, presents a novel solution: the design and
implementation of a hybrid visual-textual language, \textsc{Hybrid Clojure\-Script}, that
is both extensible and preserves static reasoning. The key feature of
the language is that programmers can construct code with a mix of
textual and \textit{visual interactive} syntax, or VIsx
(pronounced "vizzix"). As its name suggests, VIsx is not a mere
visualization of text but is also interactive. That is, editing
VIsx via clicks or other GUI gestures results in changes to the
code. At the same time, VIsx is syntax that has a textual
representation that is always available for editing, e.g. in a
plain-text editor, and its GUI will be updated according to any
changes to this text.

Further, each VIsx is an extension of the hybrid language, which
enables creating custom constructs that intuitively express concepts
from specific problem domains. The extensions are "linguistic" which
means that they become a proper part of the language and do not
require external tools or have special status in the language. Such
extensions also compose with other syntax and language constructs
including abstraction mechanisms. They can be defined and used in the
same file as other definitions, even to define other VIsx, and
can also be packaged into a library to be shared and used by other
programmers. They can also abstract over many different parts of a
program, not just expressions or statements. Finally, because the
extensions are linguistic, existing refactoring and analysis tools
continue to work for programs that use visual code elements.

The kind of extensibility described here has been mainly pioneered for
the past few decades by the Scheme macro
system~\cite{dybvig-syntactic} and the Racket language's ability to
create towers of languages~\cite{fffkbmt:programmable, acf:super}, but
such abstraction capabilities are becoming more mainstream as seen in
the macro systems of modern languages such as Rhombus~\cite{rhombus},
Rust~\cite{rustmacros}, Scala 3~\cite{scala3macros},
Julia~\cite{juliamacros}, and Elixir~\cite{elixirmacros}. A unique
feature of such a system is that expansion of a macro can produce more
macro invocations, a macro definition, or even series of macro,
function, and other definitions.

A contribution of our work is extending this "tower" to the
code-editing phase in the form of graphical syntax
extensions. Specifically, we created \textsc{Hybrid Clojure\-Script}, which adds graphical
syntax capabilities to the extensible language, Clojure, where each
VIsx is itself a proper language extension.  Having this
capability means that the result of VIsx expansion can be other VIsx
definitions. This enables many use cases, such as the meta-VIsx in
section~\ref{sub:meta}, that are not possible with other graphical
syntax systems.

A complicating issue is that arbitrary code may be run during each
expansion phase to produce the output of that phase. This arbitrary
code can even be defined by one part of a macro output, and then used
by another part.
To guarantee that definitions---including generated macro and function
definitions---are available at the proper time during expansion, it
has been shown~\cite{flatt-composable2002} that code execution should
be split into distinct code phases, e.g., "edit time" (when VIsx
code runs) vs "compile time" vs "run time". For example, code that is
used only during expansion should not be run again at run time. And a
macro that relies on another macro needs to be expanded after the
latter. With the possibility of macro-generating-macros, the separate
phases are necessary to limit how much dynamic information a
VIsx instance has access to, in order to preserve certain static
properties such as binding and hygiene, as well as enable
composability of extensions.

In contrast, other extensible systems (both graphical and
non-graphical) often only allow expansion in limited program contexts,
mostly cannot emit other definitions, and typically only have a fixed
number of expansion phases (usually one). This means that they cannot
implement all of the examples from our evaluation.

Our focus on proper linguistic extensions, and the distinct code
phases that they require, is mainly what distinguishes our work from
other graphical syntax efforts. In the criteria of
\citet{horowitz2023liverichcomposable}--- who recommend that
non-textual programming systems should be "live, rich, and
composable"---\textsc{Hybrid Clojure\-Script} prioritizes "richness", i.e., the ability to
create domain-specific graphical syntax, but as we described the
extensions go beyond what other projects are capable of. The tradeoff
for this design choice is that, while our "rich" extensions are
naturally "composable", "liveness" becomes more complex. More
specifically, maintaining separate code phases means that not all
VIsx are live. Instead, a programmer may intentionally to limit
how much dynamic information an extension can access, in order to
preserve more static reasoning. A programmer can also write VIsx
with a live feedback loop, however, and thus with \textsc{Hybrid Clojure\-Script}
programmers may create hybrid programs that exist anywhere in the
spectrum from completely static to fully live.

\subsection{Contributions and relationship to previous publication}

We used our experience with a first hybrid language
prototype in Racket~\cite{plt-tr1} to implement a second
one---\textsc{Hybrid Clojure\-Script}---that is based on
ClojureScript~\cite{m:clojurescript}, a functional language with some
commercial usage. In addition to enabling further evaluation of our
prior insights and ideas in a new ecosystem, \emph{the second
implementation contributes an additional and critical intellectual
insight} for how future hybrid languages should be built: the
construction of a hybrid language should rely on a {\em retained\/}
graphical-user system~\cite{retainedGUI2006} to render the
visualizations, not an {\em immediate\/}
one~\cite{retained-vs-immediate}.

To give some more detail, Racket's hybrid capabilities use the
language's GUI framework~\cite{plt-tr3}, which is also the basis of
its primary IDE, DrRacket~\cite{drracket}. Concretely, the VIsx
capabilities of Hybrid Racket must rest on the canvas class, which is
an immediate GUI element~\cite{retained-vs-immediate}, meaning it
immediately draws pictures into a given drawing context at a specified
position. But in order to allow a VIsx to have a textual
representation, it cannot be immediately rendered because it does not
always have access to the drawing context. Thus, its API must operate
in retained mode. As a result, the preliminary Racket version had to
re-implement a retained version of the canvas class and all other
widgets (buttons, drop-down menus, etc.).

The implications of this for VIsx users are dire. If their
application comes with a GUI that uses Racket's ordinary widgets and
if they wish to express the same geometric ideas inside of the code as VIsx,
they have to replicate this GUI with the team's new retained widgets
(which are nowhere near as polished as the existing ones and suffer
from serious performance problems). In other words, developers must
implement the same GUI twice, with two separate libraries. The
language implementers suffer as well since they now must maintain two
separate GUI libraries and keep them in sync. In short, Racket's
graphical-user system turned out to be a major obstacle for creating a
hybrid language.

By contrast, GUIs in some languages are created and manipulated
via a DOM, a retained mode graphics system.  If there exists at least
one IDE that uses the same GUI system, a team can easily modify this
IDE so that it can display VIsx code as either text or an
embedded mini-GUI or even both next to each other, thus eliminating
the need for two separate GUI libraries. In short, it can create a
hybrid language system.

The experience of implementing a second hybrid language also contributes
another way in which hybrid languages should be evaluated.
Whereas Andersen et al.'s initial
effort focused on \emph{usefulness}---demonstrating that graphical
VIsx does indeed express some concepts better than textual
code---a hybrid language should also be \emph{adoptable}---meaning
it should preserve a programmer's typical workflow as much as
possible. A big first step towards adoptability comes from creating
a hybrid language from an existing general-purpose language which
users are already familiar with, rather than forcing users to learn a
new ad hoc special-purpose language.  A related second step is
maintaining compatibility with a programmer's familiar tools---such as
the IDE---and libraries. Programmers are notoriously fickle when it
comes to their preferred tools and do not easily switch to new ones. A
third aspect of adoptability is allowing programmers to use existing
libraries when defining and programming with VIsx,
something that the preliminary implementation in Racket did not
allow.

Along these lines, \textsc{Hybrid Clojure\-Script} is an extension of the existing
ClojureScript language. It also comes with a hybrid IDE based on
CodeMirror, whose basic editing UI is already familiar to many
programmers, particularly those who use browser-based IDEs (e.g.,
users of GitHub, CodePen, JSFiddle, etc.). Finally, as mentioned,
\textsc{Hybrid Clojure\-Script} users may create and manipulate GUI components
using the standard DOM API, which is available not only in browsers,
but also in modern IDEs such as Visual Studio Code, and even some
native operating systems like Android. This means that programmers may
also use many familiar GUI libraries for the DOM such as React,
Angular, and Vue.js. All of this dramatically improves the
adoptability of our new hybrid language.

To emphasize this point, this paper presents two evaluations. First,
it compares the new hybrid language to the original
version~\cite{abf:adding} (which includes porting all the previous
examples) and exposes the critical differences. Second, it adds a new
independent evaluation of the new language with an emphasis on
adoptability that itself has two parts:

\begin{itemize}

\item{how the language affects specific programmer workflow actions; and}

\item{how programmers can use existing libraries to implement graphical syntax.}

\end{itemize}

The end result is a programming language that is much more suited
towards helping programmers express code visually, when
appropriate. The authors urge readers familiar with the
original paper to study this new material.

\subsection{A first geometric example} \label{sec:tsuro-intro}

\noindent{}The rest of this section introduces \textsc{Hybrid Clojure\-Script}
and visual and interactive syntax (VIsx) with an example.  To
demonstrate the usefulness of VIsx, consider a software
implementation of the board game Tsuro.\footnote{https://en.wikipedia.org/wiki/Tsuro} In the game,
players take turns placing square tiles containing four path segments,
where each segment connects two sides of the square. When a tile is
placed, each player's gamepiece that is touching the new tile is moved
as far along the newly extended path as possible. If this movement
forces a gamepiece off the board, its player is eliminated. The last
surviving player wins.

\begin{wrapfigure}{r}{0.35\textwidth}
  \hspace{-24pt}
  \includegraphics[width=0.4\textwidth]{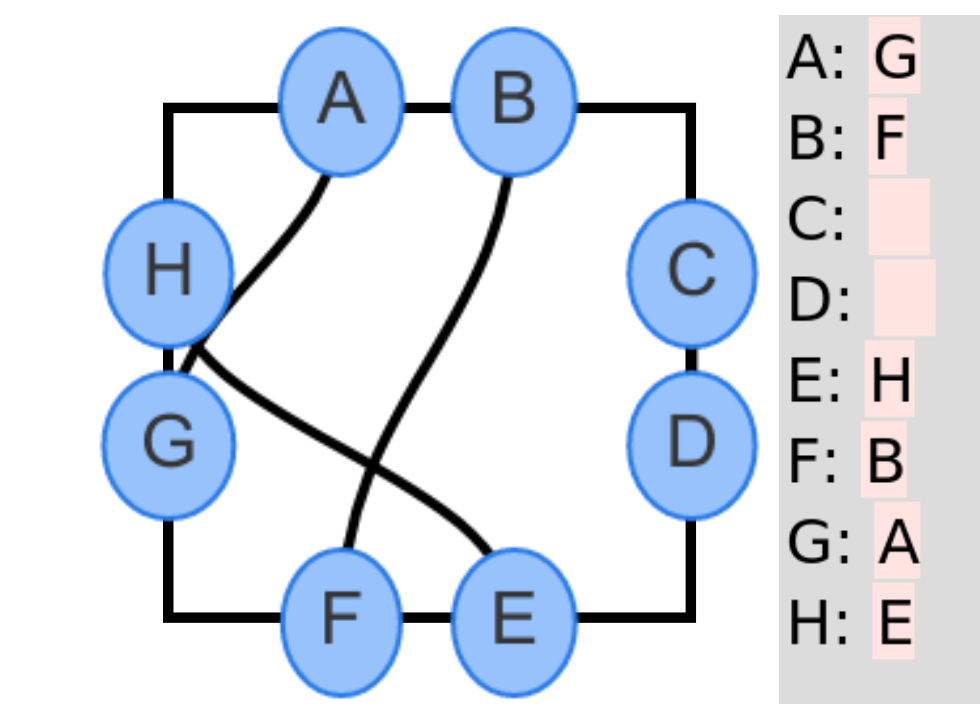}
  \caption{A Tsuro tile VIsx}
  \label{fig:tsuro-tile-ed1}
  \vspace{-24pt}
\end{wrapfigure}

Now imagine a programmer who wishes to write unit tests for this
game. If the code is written using a programming language that
supports VIsx, the tester may import a graphical representation
of the Tsuro tile code. This rendering would be automatically
displayed when the programmer is viewing the program in a
VIsx-supporting IDE. Further, a programmer can "edit" this
visual code with UI actions such as mouse clicks or menu selections.

Figure~\ref{fig:tsuro-tile-ed1} shows such a VIsx
tile, whose GUI actually consists of two different renderings: a graph
with nodes connected by paths (left) and manual entry text box fields
(right). Accordingly, programmer can "edit" this "code" in two ways:
by clicking the graphical GUI on the left or filling in the
text boxes on the right. The two sides are linked: the
graphic updates the paths whenever a user updates a text field, and
the text field updates when the programmer connects two nodes
graphically via GUI actions. In the figure, a developer is just about
to connect the nodes labeled \texttt{C} and \texttt{D}.

Each VIsx instance \emph{also has a purely textual
representation}, for three reasons. First, it allows programs with
graphical elements to be viewed and edited in any plain text
editor. This backwards compatibility makes it easier for programmers
to adopt and deploy VIsx in their programs because they will
not need to completely abandon their existing text-based
tools. Second, it means that the language does not need to be extended
with new runtime semantics for the visual elements; a VIsx is
merely an alternate (interactive) means of editing the same piece of code.
This also helps preserve programmers' tools (e.g., for
refactoring or analysis), and also saves programmers from having to
learn a completely new language. Third, it allows
textual and visual elements to be nested and interleaved in a
straightforward way, because the visual elements are part of the program's AST,
for applications that are best understood with such a hybrid rendering.

\begin{figure}[hbt]
        \centering
  \includegraphics[width=0.4\textwidth]{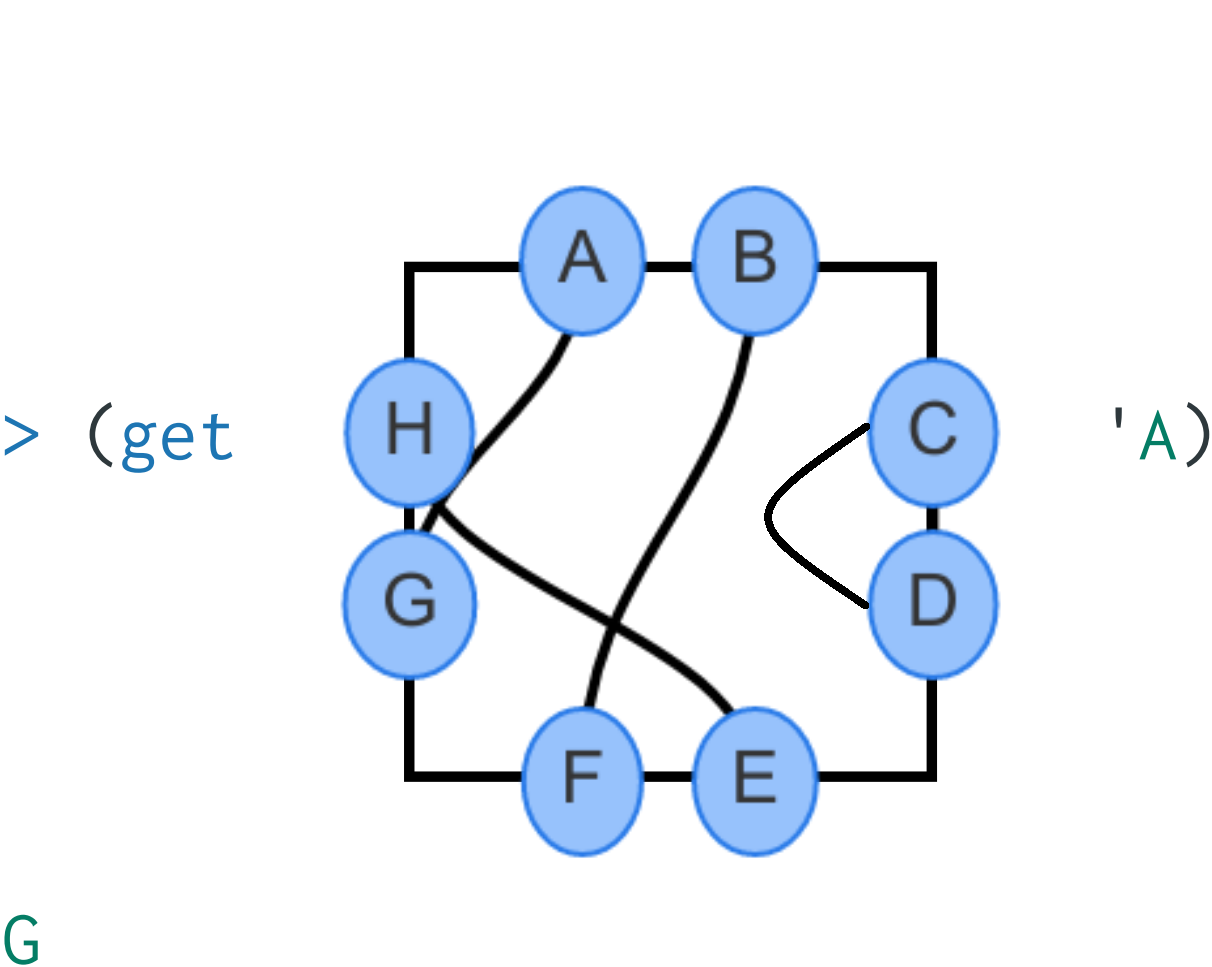}
  \hspace{20pt}
  \includegraphics[width=0.4\textwidth]{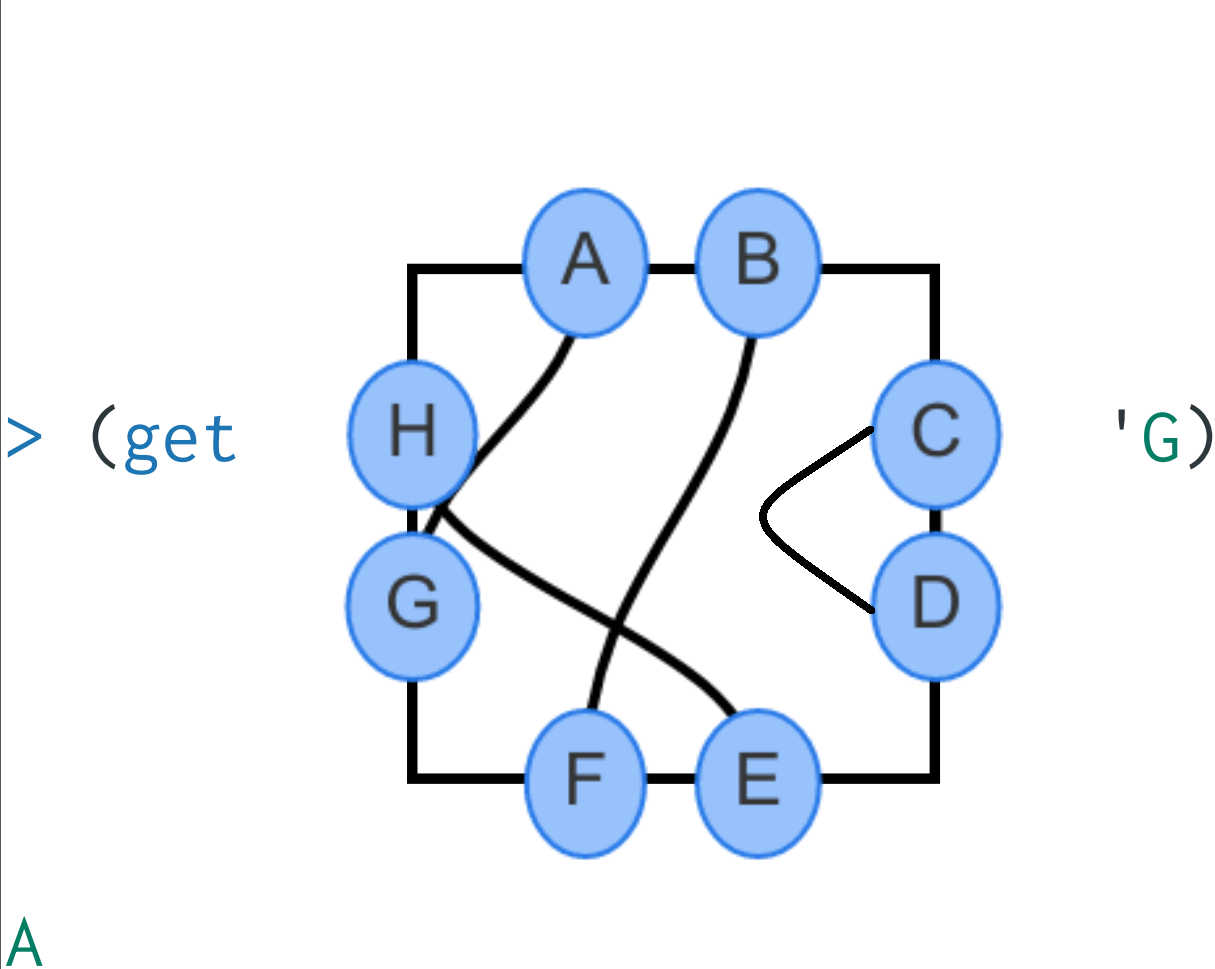}
  \caption{VIsx in a REPL}
  \label{fig:tsuro-tiles-repl}
\end{figure}

For example, figure~\ref{fig:tsuro-tile-ed1}'s Tsuro tile evaluates to
a hash-table of its connections. Thus it may be
embedded in textual code that calls hash operations, such as in
figure~\ref{fig:tsuro-tiles-repl}.

\begin{wrapfigure}{r}{1in}
\centering
\vspace{-16pt}
\hspace{-5pt}
\includegraphics[scale=0.8]{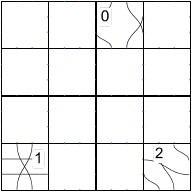}
    \caption{Board VIsx}
    \label{fig:tsuro-board}
\vspace{-24pt}
\end{wrapfigure}

A Tsuro developer could also create VIsx for the game board
itself, seen in figure~\ref{fig:tsuro-board}, which is rendered as a
grid of (slots for) tiles. In the image, three players have
each placed one tile.
Further, each tile contains a player's gamepiece represented by a
number.

\begin{figure}[hbt]
        \centering
  \includegraphics[width=\textwidth]{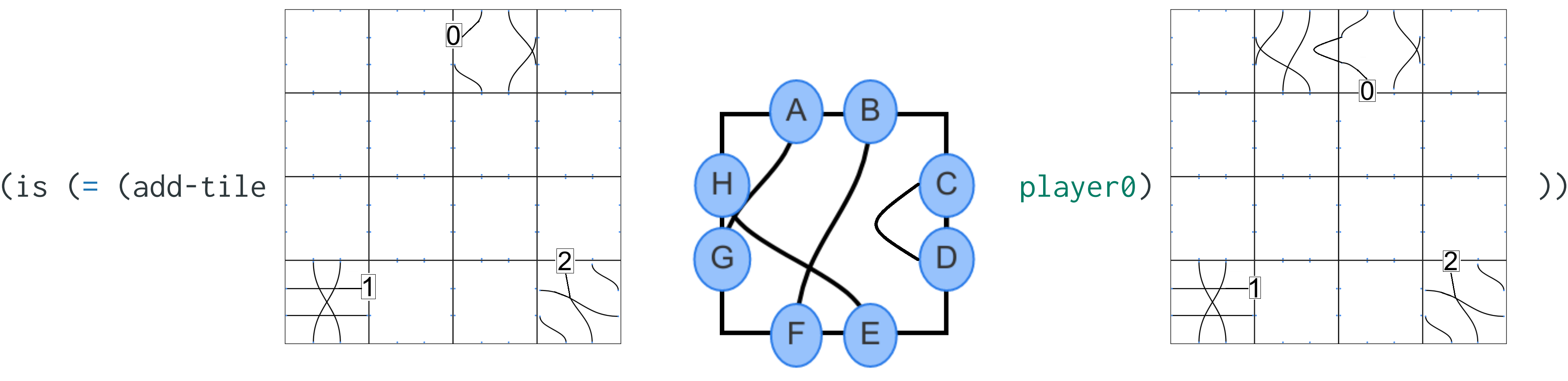}
  \caption{Unit test adding a tile to a Tsuro board, with VIsx}
  \label{fig:tsuro-add-tile-to-board}
\end{figure}

Unit test code that places a (graphical) tile onto a (graphical)
board might look as presented in figure~\ref{fig:tsuro-add-tile-to-board}.
This one-line unit test checks the \ttt{add-tile} method,
which takes three arguments: a board, a tile, and a player. Its result
is a new board state with the new tile placed touching the edge that that
player's gamepiece started on, and the player's gamepiece moved as far
as possible until it again faces an empty spot on the grid.

\begin{figure}[b]
\begin{minipage}[t]{.4\textwidth}
\begin{flushleft}
  \includegraphics[width=.85\textwidth]{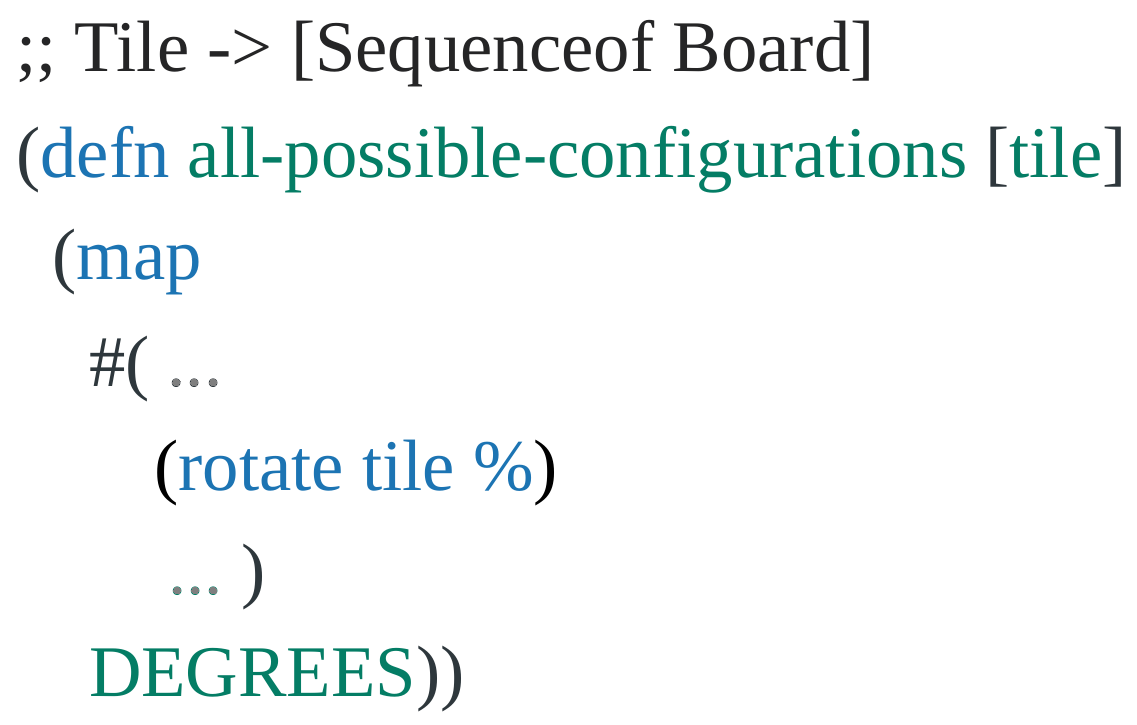}
\end{flushleft}
\end{minipage}
%
  %
\begin{minipage}{.59\textwidth}
\vspace{-16pt}
  \includegraphics[width=.79\textwidth]{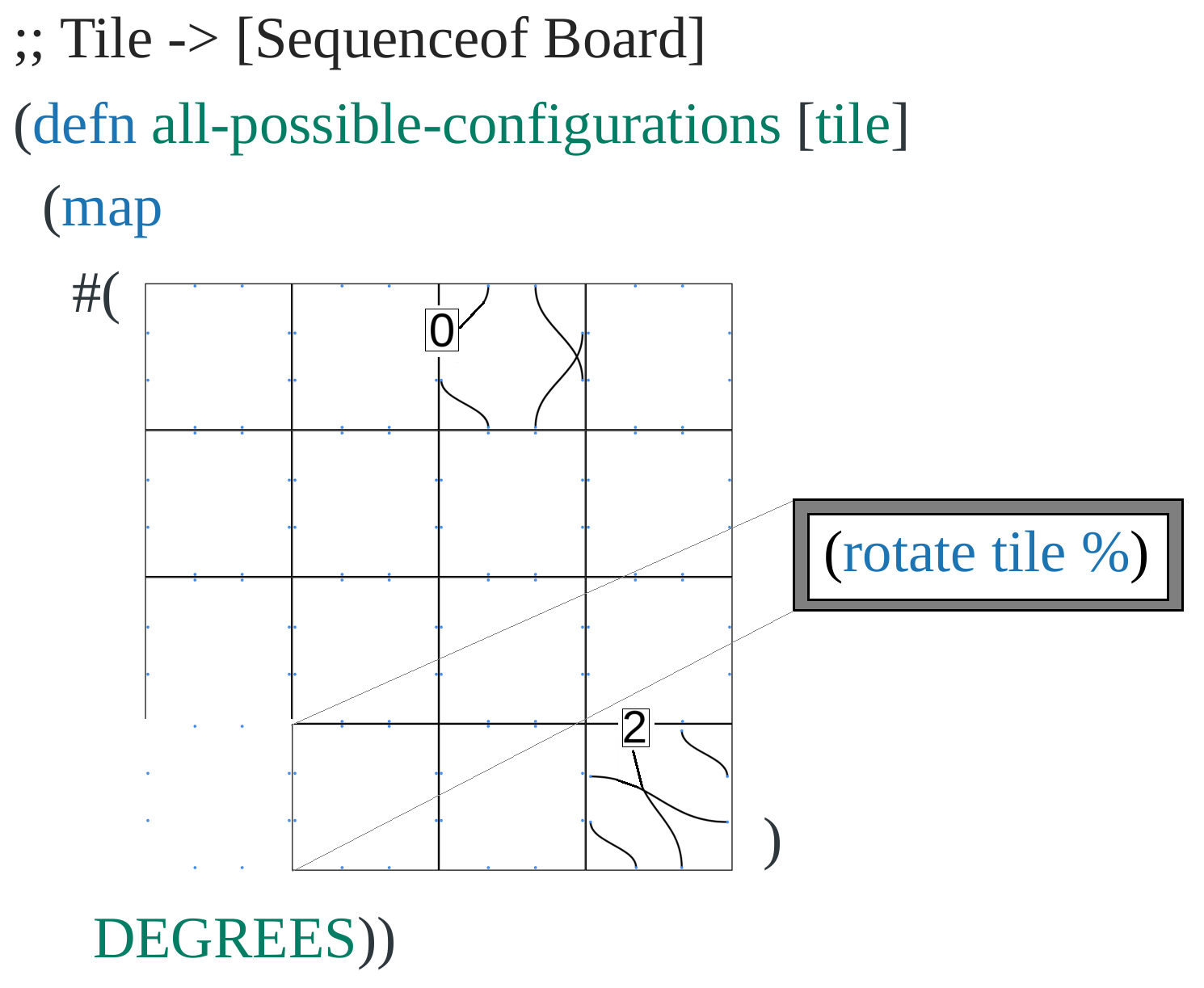}
\end{minipage}
\caption{Function placing tile at various orientations: (L) text; (R) text embedded in VIsx}
\label{fig:helper}
\end{figure}

VIsx is still syntax, however, and thus it should compose with
other syntax, both textual and visual. For example, VIsx may
appear within textual syntax (as previously described), but textual
syntax may also appear within VIsx. The Tsuro developer may
want the latter when writing a helper function for unit tests that
produces a list of board configurations for exploring moves, as seen
in figure~\ref{fig:helper}. As the (comment) signatures say, the
\ttt{all-possible-configurations} function consumes a tile and
generates a sequence of boards. Specifically, each board in the output
contains the input tile rotated by a different number of degrees. The
left side of the figure shows the purely textual code, where the
ellipses before/after the call to \ttt{rotate} represents the code
that constructs each resulting board (e.g., a call to \ttt{add-tile}
as seen in figure~\ref{fig:tsuro-add-tile-to-board}). On the right
side, the ellipses are replaced with a board VIsx. Each empty
spot on the board, where a tile is to be inserted, is itself a
VIsx, for editing textual code. Specifically, clicking a tile
pops out a one-line code editor as seen in the figure. A programmer
may write both textual code and VIsx in this nested editor, and
the nested code will have the same static properties, e.g., lexical
scope and hygiene, as the rest of the program.

\begin{figure}[t]
  \includegraphics[scale=0.8]{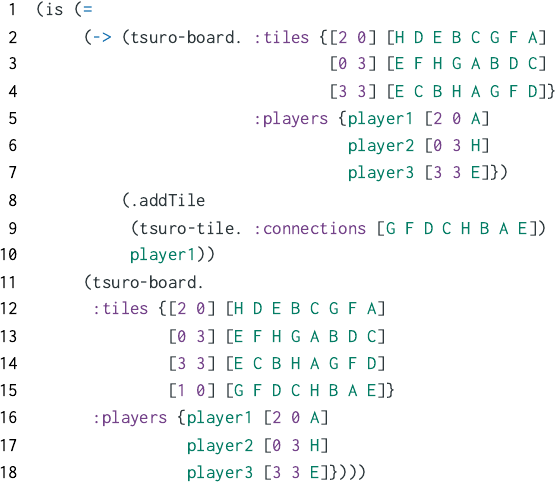}
\caption{Purely textual unit test code}
\label{tsuro-text}
\end{figure}

For comparison, figure~\ref{tsuro-text} shows the same unit test
with plain textual code. As with the graphical unit test code,
\ttt{add-tile} expects a board, tile, and player. The board is
constructed with tiles and player start locations. Each tile is a list
of eight letters, each representing its connecting nodes.
We invite the reader to improve on this textual notational choice and
compare their invention with the visual syntax in figure~\ref{fig:tsuro-add-tile-to-board}.

The Tsuro examples come from a real game implementation written by the
authors, using the presented VIsx, in \textsc{Hybrid Clojure\-Script}. This paper
itself is also written using the prototype, to help with graphical
figure manipulation. Programming in \textsc{Hybrid Clojure\-Script} is approximately
equivalent in effort to creating the textual equivalent (like in
figure~\ref{tsuro-text}) but we conjecture that most programmers
would find the hybrid code more understandable, and thus more
appropriate for expressing the problem, than text.

\section{Design space requirements} \label{sec:design-space}

While other projects share some high-level goals with ours---mainly,
allowing programmers to program with a mix of text and visual
constructs---as far as we are aware, none share the same emphasis on
linguistic extensibility, and thus they may not be able to implement
the same use cases. At the same time, our system may require certain
tradeoffs that may make it incapable of handling certain use cases
targeted by other projects. To help future language implementers who
may wish to add the capabilities of our hybrid programming system,
this section attempts to summarize some of our insights into a list of
design guidelines.

\begin{enumerate}\item A programming language should not be purely visual because linear
                       text is the dominant form of programming and will remain so for the
                       foreseeable future. Further, most code is still most clearly expressed
                       textually. Thus, developers should be able to supplement linear text
                       with visual interactive syntax only when appropriate. In other words,
                       the programming language \textbf{should be a hybrid language} where
                       textual code and visual code are mixed, on equal footing.\item At the same time, to be most useful, the hybrid language should
                                                                                       be easily "adoptable", meaning that it should preserve or improve a
                                                                                       programmer's development workflow as much as possible. A natural way
                                                                                       towards achieving this goal is to \textbf{turn an existing,
                                                                                                                                 general-purpose language into a hybrid one}.
                                                                                      As mentioned, \citet{abf:adding} did not fully achieve this because
                                                                                       programmers could not use existing GUIs and GUI libraries in their
                                                                                       implementations. Thus they had to frequently duplicate code when the
                                                                                       VIsx extensions overlapped with the GUI found in the
                                                                                       application's runtime code. \item Relatedly, the hybrid language \textbf{should be backwards
                                                                                                                                                                compatible}. This means that the hybrid language implementation can
                                                                                                                         run all existing non-hybrid code, and that existing non-hybrid
                                                                                                                         implementations can run hybrid code. An implication of this is that
                                                                                                                         visual syntax should not interfere with existing syntax---which means
                                                                                                                         that they can be nested and mixed---and that previous tools continue to
                                                                                                                         function.\item Preserving workflow implies preserving tools,
                                                                                                                                        e.g., IDEs, that programmers use with the hybrid language. More
                                                                                                                                        specifically, a \textbf{hybrid IDE should be constructed from an existing
                                                                                                                                                                one}. Since programmers have strong IDE preferences, forcing a
                                                                                                                                        different IDE choice imposes a serious adoption burden on
                                                                                                                                        programmers. Further, \textbf{hybrid code should remain
                                                                                                                                                                      backwards-compatible with existing tools}. In other words,
                                                                                                                                        programmers should still be able to view and edit hybrid code using
                                                                                                                                        non-hybrid IDEs, plain text editors, and other tools like
                                                                                                                                        source-code management.\item A hybrid language supporting visual interactive syntax
                                                                                                                                                                     \textbf{should be extensible}. Since many geometric ideas are
                                                                                                                                                                     domain-specific, programmers should be able to create and customize
                                                                                                                                                                     VIsx to best express the ideas of their problem domain. Further,
                                                                                                                                                                     the newly created VIsx extensions should be usable
                                                                                                                                                                     \textit{in any position} in a program. In our hybrid implementations, we have
                                                                                                                                                                     achieved this by starting with an extensible language that comes with
                                                                                                                                                                     an expressive macro system.
                 \item Relatedly, the visual syntax extensions \textbf{should be
                                                                       linguistic}. In other words, it should be a feature of the programming
                       language rather than, say, a third party tool that graphically renders
                       syntax or generates some code. This means that it should smoothly
                       compose with other syntax and all other language constructs, including
                       abstraction mechanisms. A VIsx definition should be able to be
                       put in a library, should be able to use other (e.g., GUI)
                       libraries, and should also be able to contain other VIsx
                       instances.\item Visual syntax extensions \textbf{should support distinct code
                                                                        phases}~\cite{flatt-composable2002}, e.g., it should have a separate
                                       edit time, compile time, and runtime. This allows preserving static
                                       reasoning in the presence of visual extensions,
                                       which increases expressiveness because programmers can create visual
                                       extensions that use static information like lexical scope. It also
                                       enables other benefits such as separate compilation, and potentially
                                       avoids other serious (e.g., security) pitfalls that can affect overly
                                       dynamic programs~\cite{evalthatmendo}.\item \textbf{Interactive syntax should be persistent.} The point of
                                                                                   visual syntax is that it permits developers to send a visual message
                                                                                   across time. Thus, in contrast to wizards and code generators, it is
                                                                                   not a GUI that merely generates textual code once. Instead, a piece of
                                                                                   VIsx should be serialized so that its relevant state is saved
                                                                                   when the program is written to file. One developer can then quit the
                                                                                   IDE, and another can open this same file later, at which point the
                                                                                   interactive-syntax editor can render itself after deserializing the
                                                                                   saved state.\end{enumerate}

Figure~\ref{fig:design-comparison} summarizes the differences, with
regard to the criteria presented, of the system presented in this
paper with the previous work of~\citet{abf:adding}. It clarifies that
\textsc{Hybrid Clojure\-Script} cooperates with IDE tools properly and is better able to
preserve the existing workflow. Further, as the following sections
explain, the edit-time GUIs use the same library as run-time GUIs and
the IDE itself. Besides making the language and IDE easier to use, it
also improves \textsc{Hybrid Clojure\-Script}'s performance compared to Andersen et al.'s
enhancement of Racket and DrRacket.

\begin{figure}[htb] \small

   \def\longestprop{\hbox{Hybrid visual and textual syntax on equal footing}}

   \newdimen\stringwidth
   \setbox0=\longestprop
   \stringwidth=\wd0

   \def\prp#1{\hbox to \stringwidth{#1}}
   \def\row#1{\hbox to \stringwidth{#1\hfil}}

   \def\comp#1{\multicolumn{1}{c}{#1}}

   \def\yes{\hspace{.8cm}yes\hspace{.1cm}}
   \def\noo{\hspace{.8cm}no\hspace{.1cm}}
   \def\yeah{\yes\relax}
   \def\nope{\noo\relax}

    \begin{tabular}{lcc}
       \toprule
       Criteria  & \citet{abf:adding} & This paper \\
\midrule
       Hybrid visual and textual syntax on equal footing      & \yeah   & \yeah \\
       Adoptable, general-purpose language (incl. GUI libs)       & \noo {\scriptsize{(Racket)}}    & \yeah {\scriptsize{(Clojure)}} \\
       Hybrid PL is backwards compatible                      & \yeah   & \yeah \\
       Adapt a popular, existing IDE                            & \noo {\scriptsize{(DrRacket)}}  & {\small{moreso}}  {\scriptsize{(CodeMirror)}} \\
       Visual syntax is extensible                            & \yeah  & \yeah \\
       Visual syntax is linguistic & \yeah   	& \yeah \\
       Visual syntax supports distinct code phases         & \yeah & \yeah \\
       Visual syntax saved with program (persistence)         & \yeah & \yeah \\
       \bottomrule
    \end{tabular}

  \caption{Comparison with~\citet{abf:adding}, according to section~\ref{sec:design-space}'s criteria.}
  \label{fig:design-comparison}
\end{figure}


\newcommand{\bezname}{Bezier}

\section{A first look at the hybrid language and IDE together}\label{sec:preview}

\begin{figure}[hbt]
  \centering
  \includegraphics[width=\textwidth]{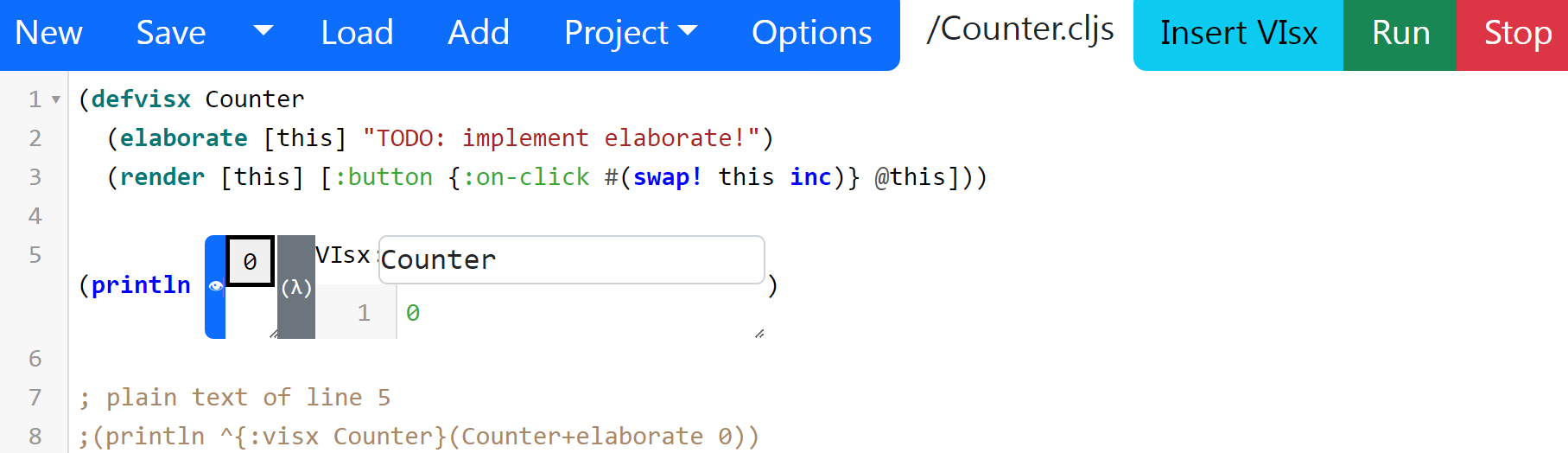}
  \caption{Hybrid IDE view of VIsx for a basic counter}
  \label{fig:counter1}
\end{figure}

\subsection{A small clickable counter in \textsc{Hybrid Clojure\-Script}}

Getting an initial sense of programming with \textsc{Hybrid Clojure\-Script} using a
hybrid companion IDE is best illustrated with a small example.
Figure~\ref{fig:counter1} shows a screenshot of a CodeMirror-based,
in-browser IDE which we call \texttt{2sIDEdMirror}, pronounced "2-sided"
for short. It contains a basic VIsx
definition---\ttt{Counter}---and its use. More specifically, the
\ttt{Counter} \ttt{defvisx} (lines 1-3) specifies two components:
\ttt{elaborate} and \ttt{render} methods (given the explanatory nature
of the example, it is presented incrementally with \ttt{elaborate}
starting as a "TODO"). Each method takes one argument named \ttt{this}
that is a mutable box (called an "atom" in Clojure/ClojureScript)
containing the state of the VISx. For \ttt{Counter}, the state is just
a number representing the current count.

To use an instance of the defined \ttt{Counter} VIsx, a
programmer can do one of two things: click the "Insert VIsx" button in
the IDE, or directly type out the VIsx as text. These
two methods produce equivalent code as lines 5 and 8 in
figure~\ref{fig:counter1} illustrate. The two lines are exactly the
same except line 8 has been commented out and thus the textual
representation of a VIsx is seen. Specifically, the commented
text on line 8 is a call to "elaborate" with the current state value,
tagged with metadata (marked by \ttt{\textasciicircum} in Clojure)
containing a reference to the \ttt{Counter} VIsx definition. If a
file containing this example were opened in a plain-text editor then
line 5 also would have this same code, except uncommented. The
existence of this purely textual representation is critical in
\textsc{Hybrid Clojure\-Script} since it allows an entire hybrid program to be saved as
just text. This means that hybrid programs remain compatible with all
existing command-line tools, plain-text editors, and other text-based
programmer tools like source control. This also means that hybrid
programs that use VIsx may still be run with non-hybrid language
versions, and thus hybrid programs are backwards compatible with all
existing language implementations.

Since the code in figure~\ref{fig:counter1} is viewed with our hybrid
IDE, however, the use of the \ttt{Counter} VIsx line 5 is
rendered visually. Specifically, there are two sides to the visual
syntax: on the right is a nested editor referencing the VIsx
name and containing the current state value; and on the left is a GUI
rendering---a clickable button---of this state value. Any update to
the state on the right causes the GUI on the left to change, and vice
versa. The GUI on the left is generated with a call to the
\ttt{render} method from the \ttt{defvisx}, which is obtained from the
metadata, along with (a mutable box containing) the current state
value as its argument. This method returns a GUI object which must
then be displayed by the hybrid IDE. For \textsc{Hybrid Clojure\-Script}, the GUI is a DOM
element---which on line 3 is constructed in the
Reagent/Hiccup-style~\cite{hiccup, reagent} using Clojure vectors
(brackets), maps (braces), and keywords (colon-prefixed
identifiers)---which our browser-based IDE knows how to display. For
\ttt{Counter}, the computed DOM element is a standard HTML button
labeled with the current count, i.e., the value represented by
\ttt{@this} (\ttt{@} is \ttt{deref} in Clojure). VIsx can
also be interactive and here the button has an "on-click" event
handler attribute which increments the count when clicked (\ttt{\#}
creates a lambda in Clojure, and \ttt{swap!} takes a mutable box and a
function and updates the box with the result of applying the function
to the current boxed value).

The \ttt{render} code that computes the VIsx GUI and handles
interaction runs at "edit time", i.e., while the programmer is writing
the code. When the programmer clicks the "Run" button, the current
VIsx state is given to the \ttt{elaborate} function, which uses
it to generate the code to be run at "run time". For the program in
figure~\ref{fig:counter1}, the run time behavior is left as a "TODO",
but this small example already illustrates how the \textsc{Hybrid Clojure\-Script}
language and its hybrid IDE must manage code execution in multiple
phases, which adds extra complication to the implementation of a
hybrid language and IDE. The next subsection discusses these different
execution phases in more detail.

\subsection{Creating a hybrid language, generally} \label{sec:design-lang}

This subsection and the one after describe more general details
about the VIsx extension mechanism, how they are realized in
\textsc{Hybrid Clojure\-Script}, and an extended \ttt{Counter} example. The first step
towards creating a hybrid language is to add a new definition form for
defining VIsx.  Such definitions must specify:

\begin{itemize}
\item{those {\em state\/} values that persist;}
\item{how a state value is {\em rendered\/} as an embedded mini-GUI in a supported IDE;}
\item{how a state value is {\em serialized\/} and deserialized as plain text, e.g., when hybrid code needs to be saved to or loaded from disk;}
\item{how the GUI {\em reacts\/} to programmer gestures which, in turn,
manipulate the state; and}
\item{how the state {\em elaborates\/} to runtime code.}
\end{itemize}

\noindent{}These requirements closely follow the model-view-controller
(MVC) architecture~\cite{mvc-dynabook, mvc-past-present2003}, which is
a common programming pattern used for creating interactive
applications, particularly those running in a web environment.

\begin{figure}[hbt]
  \centering
  \includegraphics[width=4in]{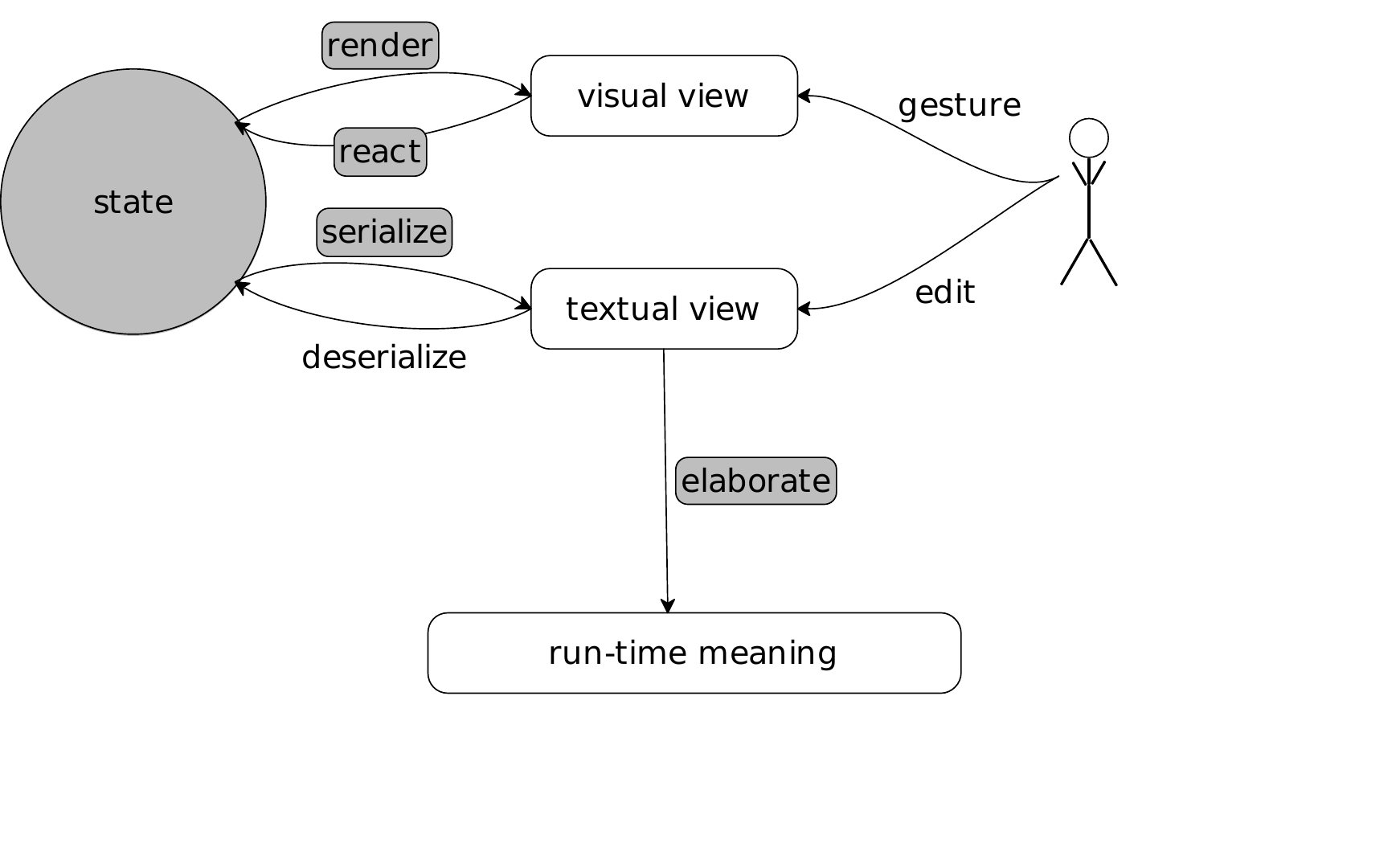}
  \vspace{-1cm}
  \caption{Interactive-syntax extensions at work} \label{fig:design}
\end{figure}

Figure~\ref{fig:design} sketches how all these components
interoperate, where the gray shapes are supplied by the VIsx
creator. In addition to the MVC parts, the compiler must also {\em
elaborate\/} VIsx instances to a runtime semantics when the
program runs.

What makes this process challenging is that the various components
execute code at different times and these execution phases must be
kept distinct from each other. This separation makes intuitive sense
because code that is only needed to render the VIsx at edit time
should ideally not be included with runtime code. Moreso, errors
occurring in one phase should not affect code running in other phases,
e.g., a program that crashes when a programmer presses "Run" should
not cause problems with VIsx GUIs, other parts of the IDE, or
the browser itself. Ideally, a hybrid IDE should allow a programmer to
gracefully control execution in all phases, e.g., with a "Stop" button
that is seen in figure~\ref{fig:counter1}. Code
should also be reusable in multiple phases if needed, e.g., a programmer
may wish to use the same GUI code at edit time---for a VIsx
GUI--- and at run time---for the main application's GUI, like in
section~\ref{sec:intro}'s Tsuro example. The uses must be distinct,
however, so that any state from the VIsx does not interfere with
the runtime application.

At the same time, a value should be able to move across phases if
necessary. For example, a VIsx state value may be required in three
phases: it is used during edit time to render the VIsx GUI, it
is used during elaboration, i.e., ``compile time'', to generate the
runtime code, and it is potentially used during the running of the
program. A language design that keeps these phases separate, yet able
to communicate, has been identified as an important characteristic of
modern extensible languages, and this design choice offers additional
benefits such as preserving reasoning about static properties like
binding~\cite{flatt-composable2002}. Managing code to run in different
phases makes the implementation of a hybrid language and IDE more
challenging, however, but section~\ref{sec:design-ide} discusses in
more detail how the \textsc{Hybrid Clojure\-Script} language and a hybrid browser work
together to enable some of these features.



\subsection{Defining VIsx in \textsc{Hybrid Clojure\-Script}}
Here is a template of a new definition form in \textsc{Hybrid Clojure\-Script}, called \ttt{defvisx}, for defining VIsx:

\smallskip
\includegraphics[scale=0.8]{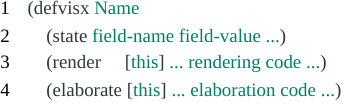}
\smallskip

\noindent{}A \ttt{defvisx} definition has the following parts, which implement the components described in the previous subsection:

\begin{itemize}

\item{A \ttt{state} specification, which is a series of field names
with initial values that will constitute the state object. Unlike the
introductory example in figure~\ref{fig:counter1}, whose state value is
a single number, most VIsx state will be an object
containing multiple fields and values.}

\item{A \ttt{render} component, which is a method with a single
  argument, named \ttt{this} by convention, that is a state value. The
  render method uses the state value fields to compute a GUI element,
  which then gets sent to the IDE, if it supports VIsx, to be
  displayed. Following standard DOM-development practice, in
  \textsc{Hybrid Clojure\-Script} \ttt{render} collapses view and control, and thus
  implements both the ``render'' and ``react'' boxes from
  figure~\ref{fig:design}. That is, it simultaneously draws the GUI
  and handles user input. For the latter, \ttt{render} may mutate the
  fields of the state and thus the current state value is received as
  a mutable box.}

\item{For \textsc{Hybrid Clojure\-Script}, programmers do not explicitly specify
serialization and deserialization because \ttt{defvisx} implements
this functionality implicitly. Specifically, state fields and values
are written as literal JSON data. Other hybrid languages without
literal syntax for data interchange formats may need to specify their
serialization more explicitly in their VIsx definition forms.}

\item{Like \ttt{render}, the \ttt{elaborate} component is a method
that consumes a single argument that is a state value. Its task
is to convert the state into runtime code, as indicated by the
``elaborate'' box in figure~\ref{fig:design}.}

\end{itemize}

\begin{figure}[hbt]
  \centering
  \includegraphics[width=4in]{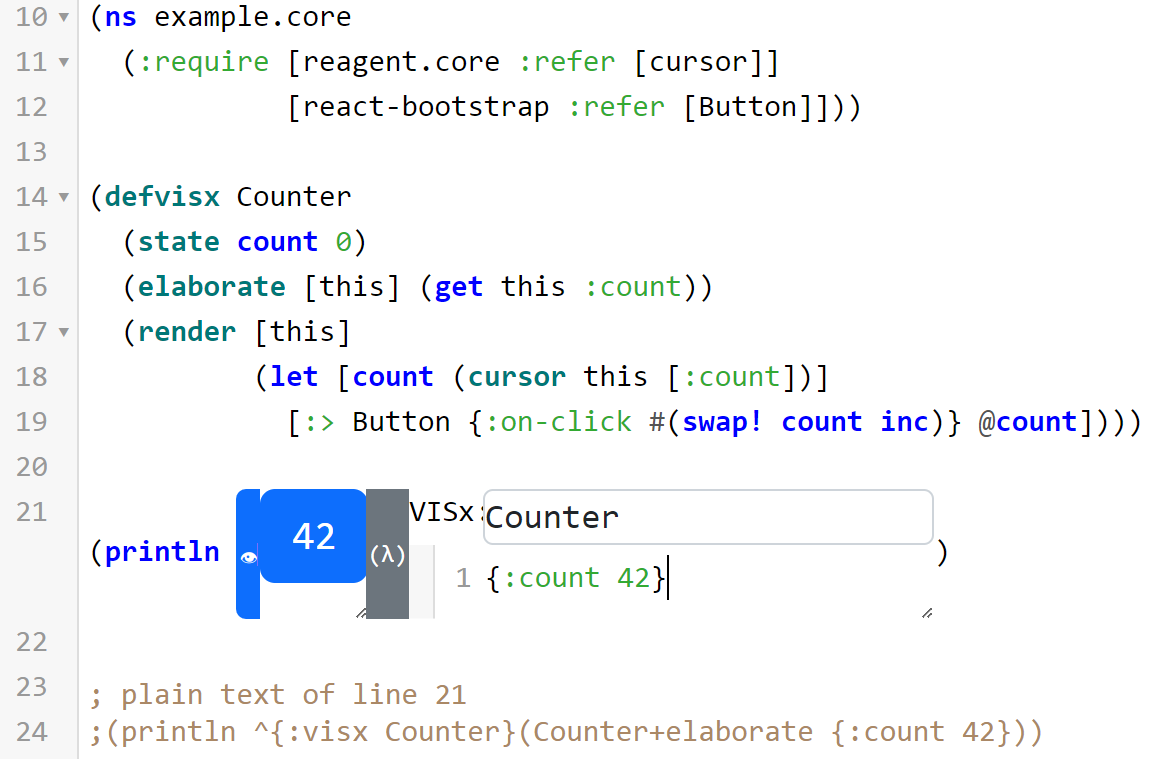}
  \caption{Extended definition of VIsx for a basic counter, now rendered with React Bootstrap}
  \label{fig:counter2}
\end{figure}

Figure~\ref{fig:counter2} presents an extended \ttt{Counter} example
that uses some of these additional \ttt{defvisx} features. The first
difference is an \ttt{ns} form that declares a namespace for
the example and imports some third-party libraries. Specifically, the
extended example now uses React Bootstrap~\cite{react-bootstrap} to
render the button, making the VIsx theme match the rest of the
IDE and, more generally, demonstrates how programmers may use
third-party GUI libraries to implement VIsx.

Another difference is that the \ttt{defvisx} has a \ttt{state}
component that specifies a field named \ttt{:count} with default value
0. An example state object with \ttt{:count} value 42 is shown both
visually as VIsx (line 21) and textually (line 24).

The \ttt{render} function still produces a clickable button as in
figure~\ref{fig:counter1}, except, as mentioned, it is now a React
\ttt{Button} instead of an HTML one (the \ttt{:>} on line 19 is a
special keyword that acts as a foreign function interface (FFI) to
external JavaScript libraries). The \ttt{render} function uses a
"cursor" (from the Reagent~\cite{reagent} library) which is a pointer
to a part of a mutable object. This makes it more concise to implement
the on-click handler, which mutates the count field in the state value
when the button is clicked. Finally, the
\ttt{elaborate} function now returns the current count when the
program is run.

In \textsc{Hybrid Clojure\-Script}, the \ttt{defvisx} definition form is a language
extension defined via a macro that when used, defines the VIsx
components, which consists of additional macro and function
definitions. Thus each VIsx definition is itself a linguistic
language extension that allows VIsx to be used in a nested
manner, i.e., to define other VIsx. This also means VIsx
may be used in many different program contexts, not just
expressions. The next section demonstrates this with a larger example.


\section{VIsx in non-expression positions}

\begin{figure}[hbt]
  \centering
  \includegraphics[width=\textwidth]{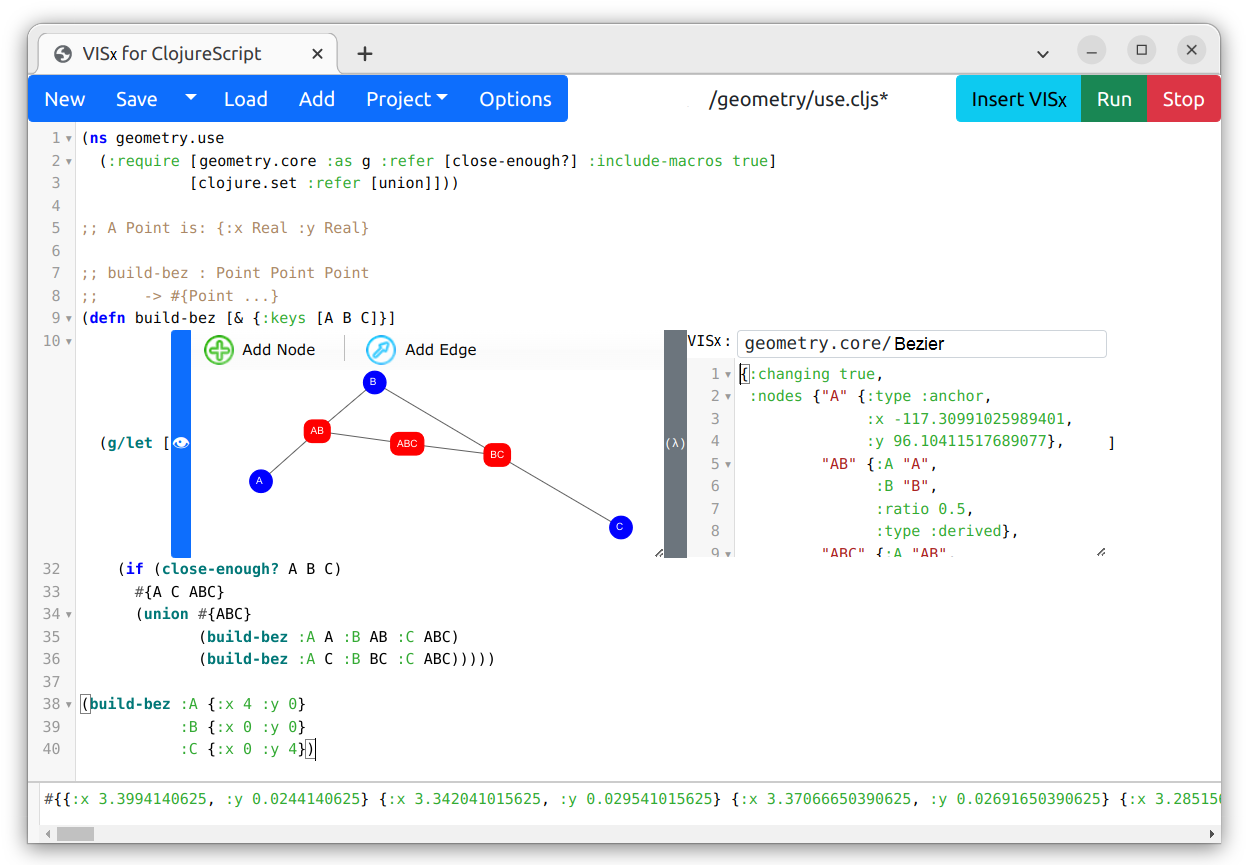}
  \caption{IDE view of the Bézier function}
  \label{fig:bez-example}
\end{figure}


This section presents another small, illustrative example showing that
VIsx may be used in non-expression positions. Specifically, it
introduces a VIsx that is a binding form.

\subsection{VIsx for a binding position}\label{sec:bez}

Figure~\ref{fig:bez-example} shows the \texttt{2sIDEd} IDE containing code
for VIsx used in a binding position. Specifically it shows a
Bézier curve function. The standard algorithm for computing a Bézier curve
combines two tasks: finding/connecting midpoints and recurring on the
first task. While the second is easily expressed via text, the first
is a geometric idea that could be better understood as a pictorial
representation. The \ttt{build-bez} function in the figure illustrates
how a programmer might do this with \textsc{Hybrid Clojure\-Script}. The function
computes a set of Bézier points from three input points---named \texttt{A},
\texttt{B}, and \texttt{C}---that form a triangle. A VIsx instance
expresses the calculation of the next three curve points---named
\texttt{AB}, \texttt{BC}, and \texttt{ABC}---graphically. Specifically, it
shows that \texttt{AB} is the midpoint of \texttt{A} and \texttt{B},
\texttt{BC} is the midpoint of \texttt{B} and \texttt{C}, and \texttt{ABC} is
the midpoint of \texttt{AB} and \texttt{BC}, matching what a student might
remember from geometry class. Since the function's input points are
unknown, the diagram does not draw a concrete figure; what is
important is the relative position of the points to each other, e.g.,
that the \texttt{AB} point is halfway between \texttt{A} and \texttt{B}.

Of course, while the graphical code effectively conveys the
program's high-level behavior at a quick glance, it actually does a
poor job of conveying some precise details, e.g., whether \texttt{AB} is
indeed exactly halfway between \texttt{A} and \texttt{B}. To check the
exact position of the nodes, we can look at the right side of the
VIsx: in the state value, the \texttt{:ratio 0.5} is what
specifies that \texttt{AB} will indeed be the midpoint of \texttt{A} and
\texttt{B}. Further, the visual and textual views are linked; a change
to either is immediately reflected in the other one, following a
standard model-view-control pattern~\cite{mvc-dynabook,
  mvc-past-present2003}. Thus, if the \texttt{:ratio} in the example
were changed---to \texttt{0.8}, for instance---then the \texttt{AB} point
might move closer to \texttt{B} in the graphic representation on the
left, and the function would compute a different kind of curve.

While a real programmer might not actually implement a Bézier function
in this way, the example serves to illustrate an interesting use of
VIsx, namely that \texttt{AB}, \texttt{BC}, and \texttt{ABC} are
binding positions (where \texttt{g/let} abbreviates a Clojure
"destructuring" \texttt{let}) for identifiers that may be referenced
later. Even though the definition of the computed point variables
\texttt{AB}, \texttt{BC}, and \texttt{ABC} are presented visually as
VIsx, the second part of the function
references them textually to compute the rest of the curve. Further, the
VIsx understands that \texttt{AB}, \texttt{BC}, and \texttt{ABC} are
binding positions, and so hybrid programs remain compatible with
existing tools, e.g., a refactoring tool that uses lexical scope
information. Preserving this kind of static information to aid program
reasoning is a key goal that differentiates our project from some
other projects that may use visual code more dynamically.

\begin{figure}[hbt]
\centering
\includegraphics[scale=0.8]{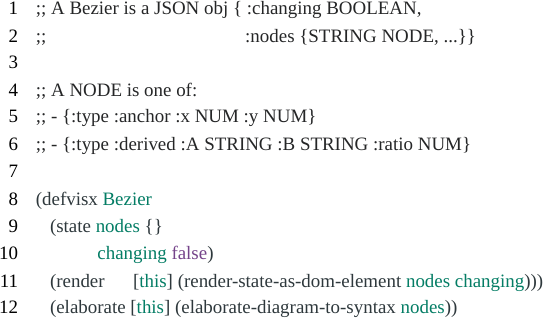}

\caption{The {\tt defvisx} for the VIsx extension in figure~\ref{fig:bez-example}}
\label{fig:the-defvisr}

\end{figure}

\subsection{Implementation of a binding position VIsx}

Figure~\ref{fig:the-defvisr} shows what a \ttt{defvisx} definition for
the midpoint extension used in figure~\ref{fig:bez-example}, called
\ttt{\bezname}, might look like.  As the comments explain, the
state consists of some \ttt{:nodes} and a boolean flag, called
\ttt{:changing}. The \ttt{:changing} field simply indicates when the
programmer is actively modifying the diagram; it is only used at
edit-time. The \ttt{:nodes} information is used for drawing the
diagram at edit time, and for setting up variable bindings in the
runtime code.

Each node can be one of two kinds: \ttt{:anchor} nodes have positions
that only become known at run time; and \ttt{:derived} nodes are outputs
of the midpoint calculation, whose values are determined algebraically
from \ttt{:anchor} nodes and other \ttt{:derived nodes}. In the
example from figure~\ref{fig:bez-example},
\ttt{A}, \ttt{B}, and \ttt{C} are anchor nodes, and \ttt{AB},
\ttt{BC}, and \ttt{ABC} are derived nodes.



\begin{figure}[hbt]
   \begin{subfigure}[c]{.5\textwidth}
     \centering
     \includegraphics[scale=0.8]{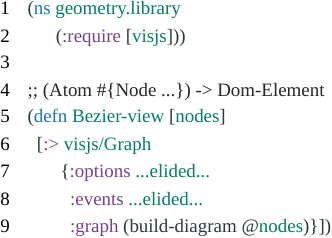}
     \caption{\texttt{library.cljs}}
   \end{subfigure}%
   \begin{subfigure}[c]{.5\textwidth}
     \centering
     \includegraphics[scale=0.8]{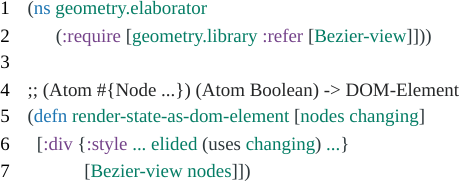}
     \caption{\texttt{renderer.cljs}}
   \end{subfigure}
   \caption{Renderer (i.e., "view" and state change "controller") for a Bézier extension}
   \label{fig:cljs-bez-render}
 \end{figure}

Figure~\ref{fig:cljs-bez-render} sketches the renderer for the
\ttt{\bezname} VIsx. Specifically, the right side defines
\ttt{render-state-as-dom-element}, which is the function given as the
\ttt{\bezname} \ttt{render} function in
figure~\ref{fig:the-defvisr}. This function will be applied to (a
mutable atom that contains) the state (which is actually passed as
separate mutable variables for each field). From this state, the
renderer computes (a nested-vector data structure that encodes) the
user-facing DOM-element: keywords in the first position of a vector
directly represent DOM tags (e.g. \ttt{:div} on line 6 represents a
\ttt{<div>} element), while functions (e.g. \ttt{\bezname-view} on
line 7) compute DOM elements dynamically.

The left side of the figure defines the \ttt{\bezname-view} function,
which draws the main \ttt{\bezname} GUI view. Conveniently, it calls
an external JavaScript library, \ttt{visjs}, to handle the low-level
drawing (the \ttt{:>} (line~6 on the left) is a special keyword
that acts as a foreign function interface (FFI) to external JavaScript
libraries).

\begin{figure}[hbt]
\end{figure}

The \ttt{visjs} library also handles programmer interaction with the
GUI, reading and modifying the state as necessary. Specifically,
\ttt{nodes} is unboxed using the \ttt{@} operator
(figure~\ref{fig:cljs-bez-render} left, line 9) and passed to
\ttt{build-diagram}, which watches with a publish-subscribe
protocol. When the state changes, the watcher notices and updates the
view.

\begin{figure}[hbt]
  \flushleft
\hspace{40pt}  \begin{subfigure}[c]{0.5\textwidth}
    \includegraphics[scale=0.8]{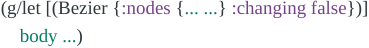}
    \end{subfigure}

    \vspace{10pt}
    \begin{subfigure}[c]{0.5\textwidth}
      \centering
      $\Longrightarrow$ \\
      elaborates
    \end{subfigure}

    \vspace{10pt}
\hspace{40pt}    \begin{subfigure}[c]{0.5\textwidth}
      \includegraphics[scale=0.8]{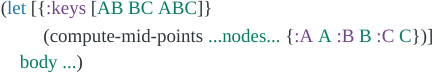}
  \end{subfigure}
  \caption{Elaboration of a \texttt{\bezname} VIsx into \texttt{let} bindings.}
  \label{fig:cljs-bez-elab}
\end{figure}

Figure~\ref{fig:cljs-bez-elab} sketches the elaboration of the (text
representation of the) Bézier VIsx to runtime code. The
\ttt{\bezname} elaborator, when applied to a state object, calls the
\ttt{elaborate-diagram-to-syntax} function (whose implementation can
be deduced from the figure and is thus not shown) that is specified in
the \ttt{elaborate} component in figure~\ref{fig:the-defvisr}. As
expected, the VIsx elaborates to a textual call of the
\ttt{compute-mid-points} function, which will compute the next three
points on the curve as explained in section~\ref{sec:bez}. The
elaboration also produces the \ttt{\{:keys [AB BC ABC]\}} binding
positions, as previously mentioned, which become part of a Clojure
"destructuring" \ttt{let} (which appears when the \ttt{g/let}
abbreviation is itself elaborated).

\section{Enhancing an IDE to support visual-interactive syntax} \label{sec:design-ide}

A hybrid programming language must come with at least one IDE that can
display VIsx extensions visually and textually. This section
explains how one might turn an existing IDE into a hybrid
one. Starting with a familiar IDE UI is critical to achieving a {\em
adoptable} hybrid language, since most programmers have strong
preferences in this regard and are unlikely to switch to a new
one. CodeMirror is a DOM-based editor that serves as the foundation of
a number of IDEs and thus its basic editing interface is already familiar
to a large number of programmers. This section describes how to equip
CodeMirror with support for \textsc{Hybrid Clojure\-Script}; the adapted IDE is called
\texttt{2sIDEdMirror}, or \texttt{2sIDEd} for short.



Since hybrid programs are still saved to files as text, a simple first
step towards creating a hybrid IDE is to mark VIsx instances
textually. As seen in figure~\ref{fig:counter1} and others, this is
achieved with a \ttt{\^{}:visx} Clojure metadata tag. This tag tells
\texttt{2sIDEd} which VIsx definition to use to visually render the
VIsx. This use of metadata plays a key role in getting an
adapted IDE to work with hybrid syntax.



After the IDE recognizes a VIsx instance, it must look up its
definition and run the code that renders the GUI. This (user-defined)
code must run at \textit{edit time}, however, which raises a number of
concerns. The most critical is that, since a VIsx GUI is in
essence a part of the IDE's own GUI, a hybrid IDE must ensure that the
edit time code that draws the VIsx GUI does not interfere with
the integrity of the IDE itself. For example, erroring VIsx code
should not crash the IDE itself. Also, any design must be modular so
that the hybrid language and the hybrid IDE do not become tightly
coupled.

\begin{figure}[hbt]
  \centering
  \includegraphics[width=0.95\textwidth]{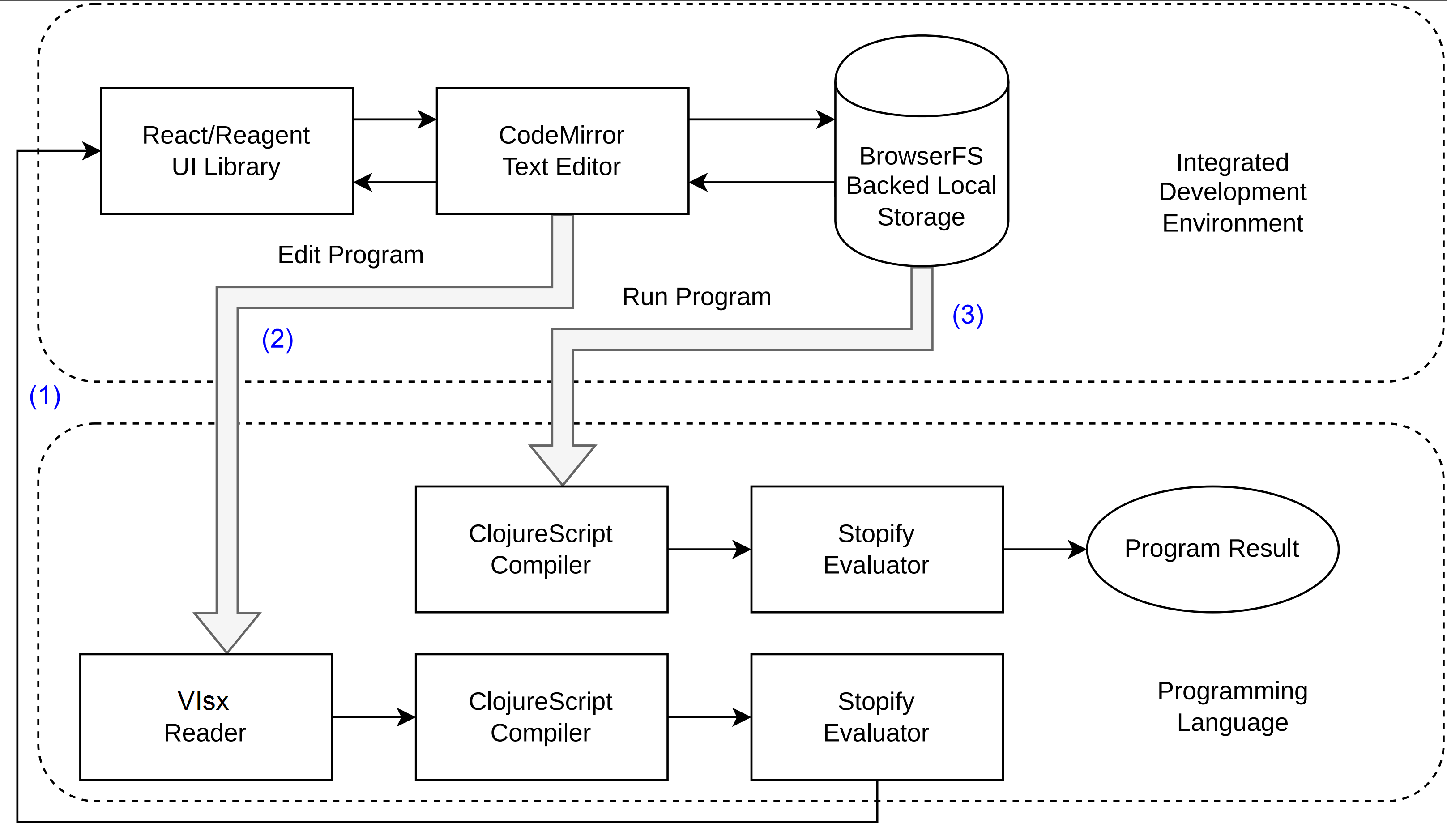}
  \caption{Architecture for \texttt{2sIDEd} (top) and \textsc{Hybrid Clojure\-Script} (bottom)}
  \label{fig:architecture}
\end{figure}

Figure~\ref{fig:architecture} shows an architecture diagram of
\texttt{2sIDEd} (top), \textsc{Hybrid Clojure\-Script} (bottom), and their interaction. It
demonstrates that \texttt{2sIDEd} supports the expected standard IDE
functionality (in the top box) such as: editing code with the
CodeMirror text editor; storing code in a file and loading it from
there;\footnote{Because \texttt{2sIDEd} runs in a browser, it uses
BrowserFS~\cite{pvb:browsix}, a lightweight file system for browsers.}
running code; and background execution.


The background execution (arrow \#2) is key to rendering the visual
syntax in a phase separate from runtime code.  As previously
mentioned, a hybrid IDE must scan the program for instances of
VIsx and run their rendering and event handler code.  This code
evaluation is handled by a combination of a ClojureScript compiler and
Stopify~\cite{bnpkg:putting}. The latter is a JavaScript transpiler
and run-time environment whose purpose is to compile plain JavaScript
into code that supports cooperative multitasking through continuation
passing.  In the context of \textsc{Hybrid Clojure\-Script}, Stopify enables two pieces
of functionality. First, it allows the IDE to pause running programs,
i.e., misbehaving VIsx extensions do not lock up the
IDE. Second, it provides a sandbox environment that separates edit
time code for VIsx extensions from the IDE. It thus prevents the
former from interfering with the IDE's internals.

When running, VIsx edit time code may need to call arbitrary GUI
libraries (arrow \#1) such as React, which also includes event
handlers. In this way, changes to the state of a VIsx instance
(as explained in section~\ref{sec:design-lang}, and illustrated in
figure~\ref{fig:cljs-bez-render}'s example) is communicated back to
the program's displayed source code.  Specifically, the rendering
code of the VIsx sends DOM elements back to the IDE to be
displayed. The code from section~\ref{sec:preview} as well as (the
right-hand side of) figure~\ref{fig:cljs-bez-render} show some
examples of how these DOM elements are computed. When these pieces of
code are sent to CodeMirror, they are placed into the text editor at
the proper places.

\section{Evaluation \#1 (adoptability): Preserving and enhancing a developer's workflow} \label{sec:evaluation}

Previous designs for visual-interactive syntax and most related work
demonstrate their {\em usefulness\/} with a plethora of examples. Each
validates that graphical syntax expresses some concepts more directly
and clearly than linear text. Evaluating a language design, however,
particularly one that is an enhancement of an existing language, should also check that it is {\em adoptable\/}.

A key element of {\em adoptability} is programmers' willingness to
use the language. One critical prerequisite is that the hybrid
language should not impose tedious tasks on programmers such as
duplication of work, should not force them to use specific IDEs, and
should not render existing tools useless. In other words, a hybrid
language is not something developers should have to learn from
scratch. Instead, they should be able to transfer what they are
already familiar with to easily create and insert VIsx
extensions.

An important dimension of adoptability is that an adoptable hybrid
language should enhance---not interfere with---the ordinary software
development workflow. These tasks include not only reading and writing
code, but also auxiliary tasks like copy-paste and searching. Many of
the latter tasks have textual-only languages in mind, but they must
continue to work with hybrid languages to ensure a smooth development
workflow for programmers. To be clear, our criterion for preservation
requires a programmer to be able to perform a chosen task with a
single chosen tool, either textual or enhanced. For example,
programmers can continue to edit code with VIsx in a plain-text
editor, and perform other operations like search-and-replace,
independently of any other IDEs that may exist. They can also do the
same tasks, even text-based ones, from solely within our enhanced
IDE. If a hybrid language requires a programmer to switch back and
forth between different editors to perform some task that could
previously be done with one editor, we consider that disruptive to the
workflow.

This section begins (section~\ref{sub:compare}) with a systematic
characterization of some major programming tasks and an analysis of
how well this paper's hybrid design compares to related projects to
enhance and preserve them. For completeness, the section concludes
(section~\ref{sub:flaws}) with an assessment of limitations of our
design. In the appendix, we further examine and
compare several more minor programmer workflow activities.

\subsection{Workflow operations and interactive syntax} \label{sub:compare}

Programmers interact with their codebases in the following \emph{major} ways:

\begin{itemize}

\item \emph{Auditing}, arguably the most common task, is reading and
  comprehending existing code. The primary goal of VIsx is to
  let code about geometric concepts speak for itself. But a secondary
  issue is preserving the static semantics of a program. If this is
  not the case, then a programmer's static reasoning about a program,
  e.g., its lexical scope, may be disrupted.

\item \emph{Creation}, another common task, is to write new code. As
 far as VIsx is concerned, ``Creation'' refers to two actions:
 (1) defining new VIsx extensions and (2) using existing
 VIsx extensions---which may be VIsx-defining meta
 extensions---to create programs. VIsx extensions should also be
 able to be packaged in, and imported from, a library, in the same
 manner that programmers are accustomed to doing with other
 abstractions. In the ideal case, a programmer working in a text-only
 IDE can still insert a VIsx and have it work correctly in a
 hybrid IDE.

\item \emph{Copy and Paste} is the act of copying code to, and pasting it from, the
 clipboard. It also refers to the direct action of dragging and dropping. Both
 are common, and visual syntax must not get in the way of either.

\item \emph{Running} programs (in the IDE or otherwise) is a
 fundamental part of software development. Existing tools should work
 without changes, even if programs include
 VIsx instances.

\item \emph{Search and Replace} is the act of finding code and, optionally,
  replacing it with new code. At a minimum, interactive syntax should not hinder
  these operations. Ideally, a developer should be able to search for graphical
  renderings of VIsx and/or replace existing code with graphical
  renderings, even if the code exists in a text editor that is nested in a GUI.

\end{itemize}

This section discusses these five \emph{major} actions from above, but there are
many more \emph{minor} actions that are nonetheless important. A
comparison of the following \emph{minor} actions, with some closely
related work, can be found in the appendix:
\emph{Abstraction},
\emph{Autocomplete},
\emph{Coaching},
\emph{Code Folding},
\emph{Comments},
\emph{Comparison},
\emph{Debugging},
\emph{Dependency Update},
\emph{Elimination},
\emph{Hyperlinking Definitions and Uses},
\emph{Merging},
\emph{Migration},
\emph{Multi-Cursor Editing},
\emph{Refactoring},
\emph{Reflow},
\emph{Styling},
\emph{Undo/Redo}.


\begin{figure}[hbt]
  \centering
  \includegraphics[width=5in]{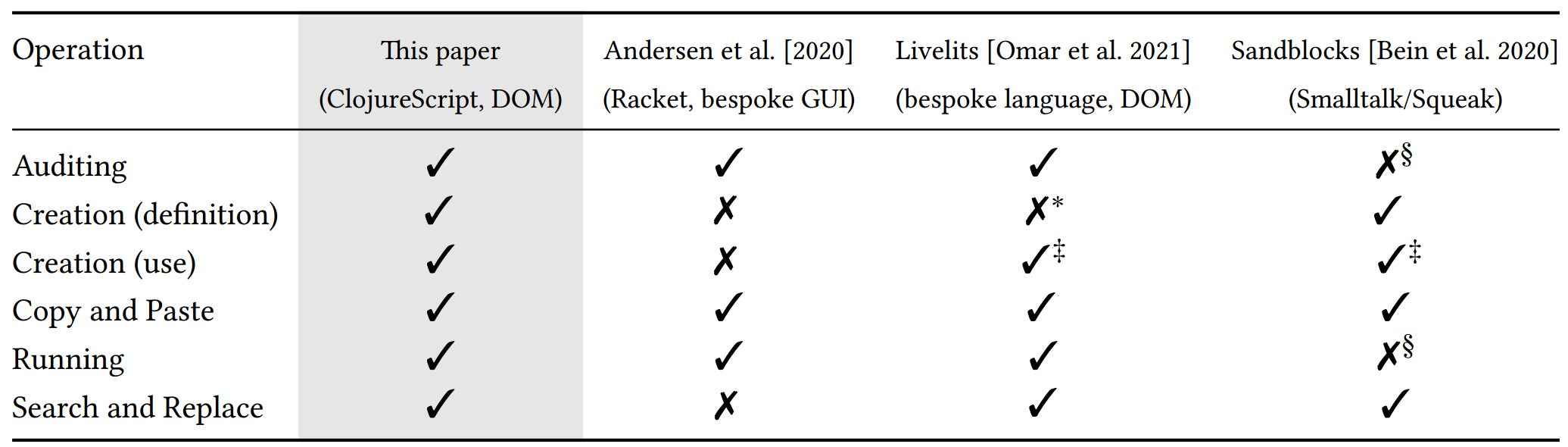}
\small
\begin{tabular}{r p{0.7\textwidth}}
    ${}^*$ & Livelit expansion cannot produce other Livelit definitions.\\
${}^\ddagger$ & Sandblocks limits interactive-syntax components to expressions; Livelits holes can appear in expressions, types, and patterns. \\
${}^\S$ & Lexical scope changed to dynamic scope, leading to possibly unexpected behavior.
\end{tabular}
  \caption{Comparison of graphical syntax and developer workflow in various systems}
  \label{fig:edit-comparison}
\end{figure}

\newcommand{\xmark}{\text{\sffamily X}}
\newcommand{\cmark}{\checkmark}


Figure~\ref{fig:edit-comparison} presents a comparison of our design
with \citet{abf:adding}, as well as two other systems:
Livelits~\cite{ocmvc:livelits, yggpmo:livepats} and
Sandblocks~\cite{sandblocks}, that are also concerned with preserving
workflow actions.  Each cell in the two columns marks a workflow
operation with a ``\cmark'' or a ``\xmark'', depending on how well the
design works with this action.




As mentioned, every VIsx is always available as both a graphical
widget and plain-text, and this is the key reason why all operations
are possible with our approach to visual-interactive syntax. Just as
importantly, the serialization works {\em inside\/} the IDE, which
means that a programmer can continue to perform traditionally
text-based actions, e.g., ``Search and Replace'', on VIsx in an
enhanced IDE, even if the text is in a code box nested inside a
VIsx GUI. Furthermore, all VIsx instances are serialized
to files as text (annotated with metadata). Hence every workflow
operation on plain-text, both inside and outside of the IDE, continues
to function.

The bespoke GUI library used in Andersen et al. is the key reason why
three major workflow operations are difficult to impossible in that
context. Specifically, the DrRacket IDE must internally save the
bespoke GUI code as binary data---rendering some existing workflow
operations, like ``Search and Replace'', unavailable. For ``Creation''
(both definition and use), the bespoke GUI library does make it
possible to define new types of interactive-syntax extensions, but
using those new extensions is challenging because developers must
leave the IDE to make even simple changes.


The Livelits project~\cite{ocmvc:livelits} aims to let programmers
embed graphical syntax into their code, similar to
\textsc{Hybrid Clojure\-Script}. Though their focus is on typed languages and using typed
holes to enable the graphical syntax, they have put considerable
effort towards preserving a programmer's workflow actions. More
specifically, they also use a textual representation to achieve this
successfully, as indicated in the figure.

Another key difference is that the Livelits work considered only an
expression language (and later allowed use in pattern
positions~\cite{yggpmo:livepats}) so there was no notion of a
"definition". A livelit, however, could be used to define the various
components of a fixed livelit definition, e.g. its view or expand
functions. If additional features were added to the language, e.g.,
modules, then livelits could generate modules in their
expansions. Incorporating such features, however, could require extending
the current single-phase expansion algorithm to include multiple
phases.

The Sandblocks system~\cite{sandblocks} adds a form of visual-interactive
syntax to the Squeak implementation of the Smalltalk programming
language, and its Morphic graphical development environment. At first
glance, their use of a dynamic language looks similar to our approach,
and their visual syntax extensions can be interleaved with program
text. Further, they are also focused on preserving a developer's tool
chain and workflow. Unlike the hybrid language presented in this
paper, however, visual elements in Sandblocks are not general; they are
limited to expressions only. For example, programmers cannot add
visualizations for field definitions, methods, patterns, templates,
and other syntactic forms. Further, the visual constructs do not
respect the language's static semantics, e.g., lexical scope. More
specifically, Sandblocks does not have a distinct edit time code
phase, separate from runtime, and thus lexically scoped variables
become dynamically scoped when used in visual extension. As a result,
the developer's toolchain and workflow are not preserved semantically.

\subsection{Minor limitations} \label{sub:flaws}

\textsc{Hybrid Clojure\-Script} is not without shortcomings. This subsection describes
three adoptability shortcomings, so that potential hybrid language
implementers are aware of them. None of these shortcomings are
fundamental to VIsx extensions, however, nor are they
fundamental to the presented design. Rather, they are trade-offs that
come with the chosen programming language, ClojureScript.

First, ClojureScript's macro system requires putting most macro
definitions and uses in separate files. This introduces some workflow
friction for programmers who wish to develop extensions and test
instances in a single file.  Fortunately, VIsx definitions and
uses can be placed in one file if they compile directly to a single
run-time function.

Second, ClojureScript's macro system is sufficiently
hygienic~\cite{kffd:hygiene, cr:mtw} so that macro-introduced bindings
never capture existing variable-references. Thus VIsx
elaborators, due to its text-based representation, inherit this same
capture-avoidance property, as seen in the \bezname{} example from
figure~\ref{fig:bez-example}, and in later examples such as the
red-black trees in section~\ref{sub:rbt}.  Because \textsc{Hybrid Clojure\-Script} uses
Clojure(Script)'s macro system, however, it is technically only weakly
hygienic, meaning capture of macro introduced variable references (but
not binders) is technically possible. But this is not an issue in
practice since ClojureScript programmers almost always use namespaces,
which resolves any potential ambiguity. Thus, variable
capture is not possible for all of the examples presented in this paper.
Finally we reiterate that this weakness in the extension system comes from
the underlying language; the VIsx extension system preserves
whatever static binding information is in those programs, and thus a
programmer does not lose any static reasoning capability.


Finally, as briefly mentioned in section~\ref{sec:design-ide},
\textsc{Hybrid Clojure\-Script} falls short in its sandboxing capabilities. By this, we
mean a guaranteed separation of edit time, compile time, and runtime
bindings, and the impossibility of edit time code to crash
the IDE it runs in. While this has not posed any problem on
the preliminary user studies we have conducted on a prototype, it does
highlight the need for future research on proper collaboration between
visual syntax and the IDE. As is, the limited sandbox provided by
Stopify means VIsx extensions are, at worst, no worse than
ordinary web pages running unsandboxed apps.

Although these limitations are undesirable, none of them reduce the
usefulness or adoptability of \textsc{Hybrid Clojure\-Script} in a substantial way. In
practice, the ability to use a rendering engine with multiple decades
of engineering offsets the high friction of defining extensions and
minor problems with hygiene and sandboxing. 
\section{Evaluation \#2a (usefulness): Re-examination of previous case studies} \label{sec:examples}

The usefulness evaluation consists of two parts. First, this section
explains how all the examples from~\citet{abf:adding}'s old hybrid
language can be ported to \textsc{Hybrid Clojure\-Script}, demonstrating that no
capabilities of the predecessor have been lost. In particular, we
highlight the ability to define custom VIsx extensions for all
kinds of language constructs, which is something that distinguishes
our system from related ones. We also highlight any
improvements that are evident with \textsc{Hybrid Clojure\-Script}. In particular, due to
the increased ability to use libraries, all the examples take less code
and effort to write. Second, in the next section, we further explore
these new benefits with new examples.

\subsection{Survey of examples where programmers want visual-interactive syntax}

We begin by surveying the original motivations of this line of work,
which are the numerous diagrams that programmers draw in place of
code. Textbook illustrations of algorithms, pictorial illustrations in
standards such as RFCs, and ASCII diagrams in code repositories and
documentation~\cite{hayatpur-2024-ascii} are just a few examples. All of
these suggest that programmers would love to express their actual code
in the same graphical manner. The following are a series of such
examples:

\begin{itemize}

\item Every algorithms book and every tree automata
  monograph~\cite{cdgjltt:tree, clrs:algo} comes with many diagrams to
  describe \textbf{tree algorithms}. Accordingly, programmers
  frequently attempt to express such algorithms graphically using
  ASCII diagrams in comments and documentation.\footnote{https://git.musl-libc.org/cgit/musl/tree/src/search/tsearch.c?id=v1.1.21}
  These diagrams contain concrete trees and depict abstract tree
  transformations.

\item Astute programmers format \textbf{matrix manipulation} code to
  reflect literal matrices when possible.\footnote{http://www.opengl-tutorial.org/} Mathematical
  programming books depict matrices as rectangles in otherwise linear
  text \cite{fgf:ampl}.

\item A typical systems course that covers the \texttt{inode} \textbf{file system
  data structure} describes it with box-and-pointer
  diagrams.\footnote{https://www.youtube.com/watch?v=tMVj22EWg6A} Likewise, source code for these data
  structures frequently include ASCII sketches of these diagrams.

\item RFC-793 \cite{p:transmission} for \textbf{TCP} lays out the
  format of messages via a table-shaped diagram. Each row represents a
  32-bit word that is split into four 8-bit octets.

\item Many visual programming
  environments, such as Game Maker \cite{o:teaching}, allow developers
  to lay out their programs as actors placed on a spatial grid. Actors
  are depicted as pictorial avatars and the code defining each actor's
  behavior refers to other actors using avatars.  In other words,
  \textbf{pictures act as the variable names} referencing objects in this
  environment.

\item \textbf{Videoediting} is predominantly done via non-linear,
  graphical editors. Such purely graphical editors are prone to force
  people to perform manually repetitive tasks, or resort to bespoke
  DSLs~\cite{acf:super}.

\item \textbf{Circuits} are naturally described graphically. Reviewers
  might be familiar with Tikz and CircuitTikz, two LaTeX libraries for
  drawing diagrams and specifically circuit diagrams. Coding diagrams
  in these languages is rather painful, though; manipulating them
  afterwards to put them into the proper place within a paper can also
  pose challenges. Relatedly, electrical engineers code circuits in
  the domain-specific SPICE~ \cite{vhn:ngspice} simulation language or
  hardware description languages such as Xilinx ISE. While both come
  with tools to edit circuits graphically, engineers cannot mix and
  match textual and graphical part definitions.

\end{itemize}

Figure~\ref{fig:worked-examples} summarizes the efforts of evaluating
the above examples (and others from this paper). The first two columns report
lines of code, for the hybrid Racket language of~\citet{abf:adding},
and \textsc{Hybrid Clojure\-Script}, respectively. The
ClojureScript implementations are usually a fraction of their Racket
counterparts. This is due to the numerous libraries that
exist in the JavaScript ecosystem and the ability of our language to
use them. The ``libraries used'' column lists any JS libraries that were
used in each example.

\begin{figure}[h]
  \begin{center}
  \begin{tabular}{lrrl}\toprule
    Name & Racket & Clojure & Libraries Used\\ \midrule
    Tree Algorithm & 353 & 118 & Vis.js\\Matrix & 175 & 44 & React Data Grid, MathQuill\\File System Data & 178 & 34 & React Data Grid\\TCP Parser & 98 & 34 & React Data Grid\\Pictures As Bindings & 88 & 20 & -\\Video Editor & 80 & 56 & Scene.js, React Video Editor\\Circuit Editor & 307 & 158 & LogicJS\\Tsuro & 408 & 159 & Vis.js\\Form Builder & 119 & 30 & Bootstrap\\Trace Contracts & - & 292 & Vis.js\\Settlers of Catan & - & 71 & React Hexgrid\\Hexgrid Game & - & 119 & React Hexgrid\\7-Segment Display & - & 148 & -\\
  \bottomrule\end{tabular}\end{center}
  \caption{Implementation efforts for worked language extensions}
  \label{fig:worked-examples}
\end{figure}

\subsection{VIsx for different language constructs}

These examples from the previous section not only show the usefulness of
graphical syntax, they also highlight a novel feature of our particular
designs when compared to other visual-programming related work. With
our VIsx extensions, programmers can create
custom graphical syntax for \textit{any} program construct, even other
syntax extensions.

This subsection describes the various
language constructs that programmers might want to abstract over, and
also summarizes which ones are used by the previous examples (in
figure~\ref{fig:examples-to-contexts}).

\begin{itemize}

\item The simplest role is that of a \textit{data literal}.
In this role, developers interact with the syntax only to
enter plain text, which the elaborator typically translates into basic
data structures. As the Tsuro examples in the introduction point out,
data-literal forms of interactive syntax typically replace large blocks
of (difficult to read) textual code.

\item The \textit{code template}'s role generalizes data literals. Instead of
  directly writing textual code, a developer inserts code into text fields
  of a VIsx. The Tsuro board in the introduction and the tree in
  figure~\ref{rb-balance} are examples of such templates. The templates build
  a board and tree, respectively, using pattern variables embedded in a VIsx.

\item \emph{Pattern matching} is commonly found in
  functional languages like Scala and Clojure and VIsx
  can enhance its readability, as the tree example in section~\ref{sub:rbt}
  shows. In this context, a developer fills a VIsx instance with pattern
  variables, and the code generator synthesizes a pattern from the
  visual parts and these pattern variables. In this role,
  pattern-matching VIsx serve as binding constructs.

\item \textit{Control flow} interactive syntax is also straightforward, with
graphical renderings elaborating to analogous textual \textit{control
                                                              flow} code.

\item Since syntax extension is a form of meta programming,
  interactive syntax naturally plays a meta-programming role as
  well. We refer to this role as \textit{meta-binding} in
  figure~\ref{fig:examples-to-contexts}. Specifically, VIsx are
  used to construct new types of syntax, and because the prototype is
  in \textsc{Hybrid Clojure\-Script}, they can generate both graphical and textual syntax
  extensions. See section~\ref{sub:meta} for an example.

\item Finally, VIsx can play the role of (other)
  \textit{binding forms}.  This role allows multiple VIsx instances to ``talk'' to each
  other. The data structure in the file system example is an example of this.

\end{itemize}

\begin{figure}[h]
  \begin{center}
  \begin{tabular}{lcccccc}
     & \rotatebox{60}{Other Binding} & \rotatebox{60}{Template} & \rotatebox{60}{Meta Binding} & \rotatebox{60}{Pattern Matching} & \rotatebox{60}{Data Literal} & \rotatebox{60}{Control Flow}\\ \toprule
    Tree Algorithm &  & \checkmark &  & \checkmark & \checkmark & \\Matrix &  & \checkmark &  &  & \checkmark & \\File System Data & \checkmark &  &  &  & \checkmark & \\TCP Parser &  &  &  & \checkmark & \checkmark & \\Pictures As Bindings & \checkmark &  &  &  &  & \\Video Editor &  & \checkmark &  &  & \checkmark & \\Circuit Editor &  &  &  &  & \checkmark & \\Tsuro &  & \checkmark &  &  & \checkmark & \\Form Builder &  & \checkmark & \checkmark &  & \checkmark & \\Trace Contracts & \checkmark &  &  &  & \checkmark & \checkmark\\Settlers of Catan &  &  &  &  & \checkmark & \\Hexgrid Game &  &  &  &  & \checkmark & \\7-Segment Display &  &  &  & \checkmark & \checkmark & \\
  \bottomrule\end{tabular}\end{center}
  \caption{Attributes of the worked language extensions}
  \label{fig:examples-to-contexts}
\end{figure}

\subsection{In depth: Red-black trees}\label{sub:rbt}

\begin{figure}
  \includegraphics[scale=0.8]{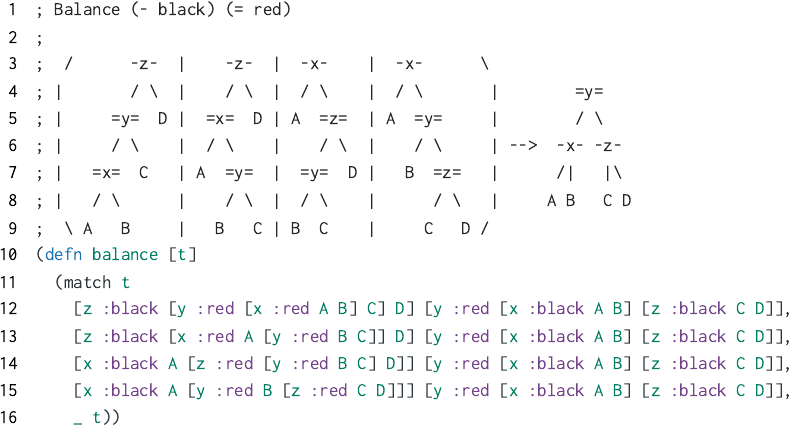}
  \caption{A textual balance function for a red-black tree}
  \label{text-balance}
\end{figure}

A key use case for VIsx is working with data structures. When
programmers explain tree algorithms, for example, they frequently
describe the essential ideas with diagrams. Often these diagrams make
it into the library documentation but then a maintainer must ensure
that the code and documentation diagrams remain in sync. An inadequate
fix is to render such diagrams as ASCII art; this at least puts the
code and diagram in the same place but still does not solve the
problem.\footnote{https://blog.regehr.org/archives/1653}

The balancing algorithm for red-black trees \cite{b:symetric}
illustrates this kind of work particularly well.
Figure~\ref{text-balance} shows a code snippet from a
tree-manipulation library in \textsc{Hybrid Clojure\-Script} . The snippet depicts a
function for balancing red-black trees using pattern
matching. The comment block (lines 1--9) makes up the
internal ASCII-art documentation of the functionality, while
the code itself (lines 10--16) is written with Clojure's
pattern-matching construct.

\begin{figure}
  \includegraphics[scale=0.8]{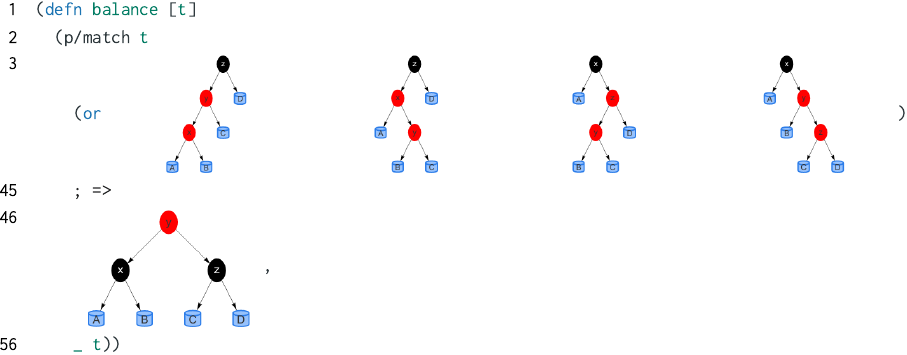}
  \caption{A hybrid balance function for a red-black tree}
  \label{rb-balance}
\end{figure}

A version of the code that uses VIsx, seen in
figure~\ref{rb-balance}, empowers developers to express the
algorithm directly as a diagram, which guarantees that the diagram and
the code are always in sync. An important and unique aspect of this
example is that VIsx can show up in the \textit{pattern} part of a
\ttt{match} expression as well as the body of the clauses, something
that many related projects do not support, and both are situated
within ordinary program text.

Concretely, in the first clause, the \ttt{or} pattern combines
four sub-patterns; if any one of them matches, the pattern matches
the given tree \ttt{t}. Each sub-pattern
represents a situation in which a rebalance is needed. The situation
may remind readers of the diagram in Okasaki's functional
implementation \cite{o:red-black}, which uses the same exact four trees on
the second page of his paper. The sub-patterns name
nodes---\ttt{x}, \ttt{y}, and \ttt{z}---and
subtrees---\ttt{A}, \ttt{B}, \ttt{C}, and \ttt{D}---with
consistent sets of pattern variables.

The clause body, or \textit{code template}---on the second line---refers
to these pieces of the pattern. It is also a VIsx and shows how
the nodes and sub-trees are put into a different position. The
resulting tree is clearly balanced relative to the matched
subtrees. Even better, since both the pattern language and VIsx
may be nested, we could allow the insertion of both text and graphical
nested patterns in each node GUI. Of course, too many nested layers of
graphical patterns may not be practical in terms of producing readable
code and our preliminary experimentation with such nested VIsx
confirmed this, so we do not show any deeply nested examples in this
usefulness evaluation.

In sum, the code consists of four input patterns that map to the same
output pattern. Any programmer who opens this file in a
VIsx-supporting IDE like \texttt{2sIDEd} will immediately understand
the relationship between the input tree shapes and the output
tree---plus the connection to the original Okasaki paper.
\subsection{In depth: Meta-extensions} \label{sub:meta}

To demonstrate that VIsx is truly a linguistic feature of a
programming language---and not some arbitrary third party add-on
tool---this section shows that it naturally cooperates with other
abstraction mechanisms in the language, including other VIsx
extensions. Specifically, we show how to create a meta-extension,
which is a VIsx that elaborates to another VIsx!

\begin{figure}[ht]
  \begin{subfigure}[b]{.5\textwidth}
    \centering
    \includegraphics[scale=0.8]{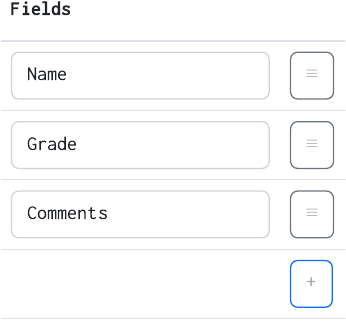}
    \caption{An assignment-specific form for graders}
    \label{fig:form-builder-def}
\end{subfigure}%
\begin{subfigure}[b]{.5\textwidth}
  \centering
  \includegraphics[scale=0.8]{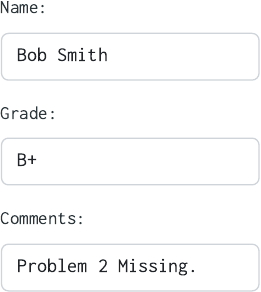}
  \caption{Instance of the assignment-specific form}
  \label{fig:form-builder-use}
\end{subfigure}
  \caption{A (meta) VIsx for a form builder}
  \label{fig:form-builder}
\end{figure}

An example of such a scenario is in the editing of (tabular)
forms. Editing forms is often self-referential, meaning a VIsx for
a form must be able to generate forms. Our concrete use case comes
from personal experience in the classroom, where a programming
instructor wants to make grading forms that teaching assistants can
use to report a student's score.

Figure~\ref{fig:form-builder} illustrates how \textsc{Hybrid Clojure\-Script} can realize
such (meta) form editors.  Specifically,
figure~\ref{fig:form-builder-def} displays a form editor for creating
grading forms. In addition to creating new fields, the instructor can
reorder fields and add optional constraints on the data stored in
those fields. The "elaborate" computation of this
VIsx creates extensions whose instances look
like the forms in figure~\ref{fig:form-builder-use}. Once filled with
student-specific data, these generated forms elaborate to
dictionaries, which can be submitted to the instructor's gradebook
code.

Our Clojure version improves on the Racket use case because the form
builder itself, as well as the forms created with that builder,
exploit another DOM-based GUI
library.\footnote{\label{note:bootstrap}https://getbootstrap.com/} Using this library, the
implementation is less than 50 lines of code. For
comparison, Andersen et al.~\cite{abf:adding} needed more than 100 lines of
code. 
\section{Evaluation \#2b (usefulness): New case studies} \label{sec:examples2}

One key improvement of \textsc{Hybrid Clojure\-Script} over Hybrid Racket is the ability to reuse
generic (GUI) libraries for the construction of VIsx. This section
illustrates this point with two examples, implemented with just \textsc{Hybrid Clojure\-Script}.
Specifically, section~\ref{sub:rest} shows how to use a graph library to express a
REST API connection as a state-machine. Similarly, section~\ref{sub:catan} uses a
general hexagon-drawing library to create a VIsx for the uniquely-shaped
board of the Settlers of Catan game.

Finally, section~\ref{sub:nested} shows how VIsx GUIs may be
nested within each other, even when mixed with textual code.

\subsection{Using and combining existing libraries} \label{sub:rest}

A finite state machine is a ubiquitous programming pattern. One common
use case is a protocol for calling API methods in a certain order,
e.g., during authentication. A state-machine also has a convenient,
well-known visual representation, which is often used in the
documentation that explains the proper method call order. In a
programming language supporting VIsx, such a diagram can
be directly added to the program as code.

%

Specifically, an API author can describe the protocol
graphically and then have the corresponding run-time checks generated
automatically. Consider an authentication protocol for a REST API on
objects with three methods: \ttt{auth}, \ttt{req}, and \ttt{done}.
The protocol imposes the following constraints on these methods:

\begin{enumerate}
  \item the \ttt{auth} method sends credentials and receives an
    authentication token in response;
  \item the \ttt{req} method, with an endpoint URL and the valid authentication token, repeatedly requests data; and
  \item the \ttt{done} method ends the authenticated session.
\end{enumerate}

\begin{figure}[ht]
  \centering
  \includegraphics[scale=0.8]{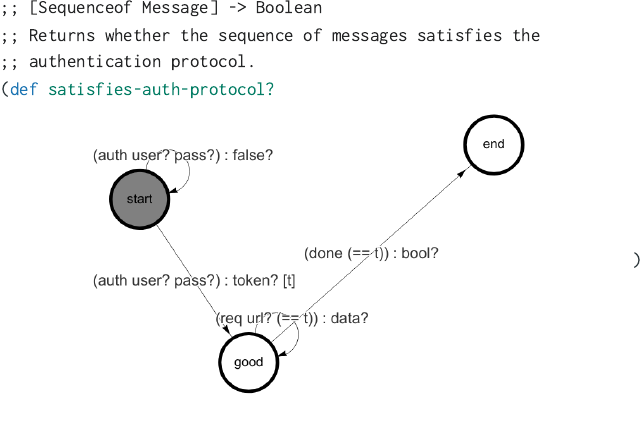}
  \caption{A state machine VIsx for an API protocol}
  \label{fig:automaton-graph}
\end{figure}

\noindent{}Figure~\ref{fig:automaton-graph} shows how a programmer may
use VIsx to express the protocol as a state machine. The state
machine has three states: \ttt{start}, \ttt{good}, and \ttt{end},
where \ttt{start} (shaded gray) is the start state.  Each state
transition is labeled with a method name and predicates for its
arguments and result. For any state, a client module can call only
methods for which there is a transition. For example, in the
\ttt{good} state, a client can call either the \ttt{req} or \ttt{done}
method.  Further, the state machine moves to a new state only if the
arguments and result of the method call satisfy the predicates on the
transition edge.  At compile time, this graphical "code" elaborates to a
predicate which, given a sequence of method calls at run time, determines whether
it satisfies the protocol~\cite{moy-trace-contracts-jfp2023}.  If a client attempts
to call a method for which there is no transition, or supplies
arguments not satisfying the transition predicates, then the protocol
is violated, and this violation is reported.

In figure~\ref{fig:automaton-graph}, the transition corresponding to a
successful authentication binds the returned token to the variable
\ttt{t}. This is shown in square brackets. The scope of this
binding includes all downstream transitions. Any transition in scope
can then use this variable in predicates.  For example, the expression
\ttt{(== t)} constructs a predicate that determines if a value is
equal to the token.

\begin{figure}[ht]
  \centering
  \includegraphics[scale=0.8]{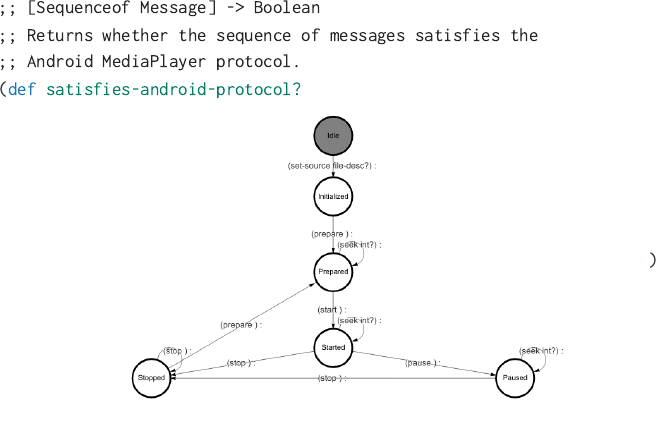}
  \caption{The Android MediaPlayer protocol using the state machine VIsx}
  \label{fig:android}
\end{figure}

Even better, the diagram presented in figure~\ref{fig:automaton-graph}
is constructed using a general-purpose VIsx for state
machines. Thus, the same code can be reused by other libraries. For example,
figure~\ref{fig:android} shows a (simplified) implementation of the
Android MediaPlayer API\footnote{https://developer.android.com/reference/android/media/MediaPlayer} protocol using the
same VIsx.

This general state diagram VIsx definition specifies
general handlers for GUI gestures to: (1) create new states; (2) delete
existing ones; (3) add or delete transitions; (4) edit the source and
destination of a transition; (5) turn states into starting or accepting
states; (6) rename states (via a text box); (7) edit the predicates labeling a
transition (via a text box); and (8) change which variables are
bound. Further, the gestures are intuitive. For example, creating a new
transition merely requires clicking and dragging from the source state
to the destination state. Altering the properties of a transition
involves selecting the transition and clicking the edit button.

The VIsx elaborator analyzes code on the transitions to determine the
necessary binding structure. Specifically, the elaborator creates a separate
function for each transition with the appropriate parameters, and provides the
run-time system enough type information to supply the correct arguments to each
function. Syntax and type errors in the specification are raised at compile
time. For example, if a transition predicate were to specify a dependency on a
variable that is not in scope, \ttt{elaborate} would signal a compile-time
error.

Overall, developing this VIsx was a low-effort project, because
it uses the \ttt{visjs} generic graph-drawing library (the same as the
Bézier curve example), and the BootStrap GUI library for creating the
transition edge editor (the same as the meta-extension in
section~\ref{sub:meta}). In total, it required writing less than 300
lines of code.
\subsection{Catan: Adapting libraries is easy} \label{sub:catan}

As illustrated with the Tsuro example in section
\ref{sec:tsuro-intro}, implementing a "board game" requires expressing
many geometric ideas and thus greatly benefits from visual code. This
subsection examines another board game example, the popular {\em
Settlers of Catan\/},\footnote{https://en.wikipedia.org/wiki/Catan} and shows how it can benefit
from VIsx, implemented with a straightforward adaptation of
off-the-shelf GUI libraries. That is, when the libraries don't provide
exactly the required functionality, it is straightforward to modify
their behavior.

Implementing {\em Settlers\/} is challenging due to its
hexagonal grid board where each edge is colored according to the
player that "owns" it. Accordingly, a {\em road\/} consists of a
continuous sequence of edges of the same color. When the game is
scored, the longest such road plays a role.

\begin{figure}[ht]
  \begin{subfigure}[c]{.52\textwidth}
    \centering
     \includegraphics[scale=0.8]{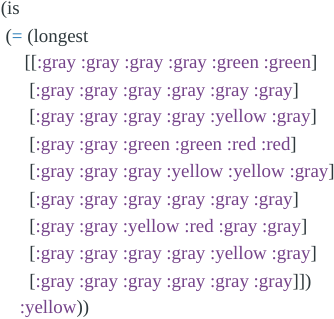}
    \caption{Textual unit test}
    \label{fig:catan-text}
  \end{subfigure}%
  \begin{subfigure}[c]{.48\textwidth}
    \centering
    \includegraphics[scale=0.8]{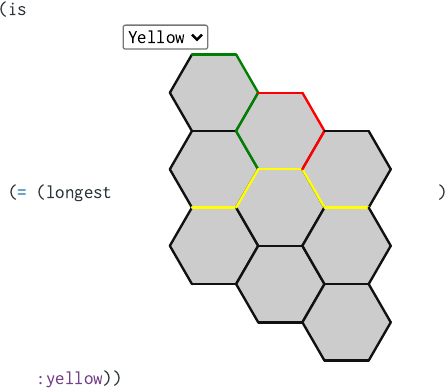}
    \caption{Visual unit test}
    \label{fig:catan-hybrid}
  \end{subfigure}
  \caption{VIsx for an implementation of ``Settlers of Catan''}
  \label{fig:catan}
\end{figure}

Unit tests again demonstrate the usefulness of VIsx extensions
particularly well. For example, figure~\ref{fig:catan-text} displays a
unit test for the longest-road calculation using traditional
plain-text syntax. In contrast, figure~\ref{fig:catan-hybrid} shows
the same unit test written with VIsx code for tiles and
boards. The right example is easier to understand because the
code renders visually. To "edit" this visual code, a programmer clicks
directly on an edge to change its color (via a drop-down menu). Even
better, this rendering is the same GUI, and reuses the same GUI code,
as the game itself, making it easy to implement. If the game's GUI
code changes, the VIsx would be updated automatically. The
game's GUI itself uses a generic open-source library for drawing
hexagons,\footnote{https://github.com/Hellenic/react-hexgrid} slightly modified (45 lines) to
enable highlighting roads. The example itself requires only about 70
lines of code.

\subsection{Multiple nested VIsx and text} \label{sub:nested}

\begin{figure}[ht]
  \centering
  \includegraphics[width=3in]{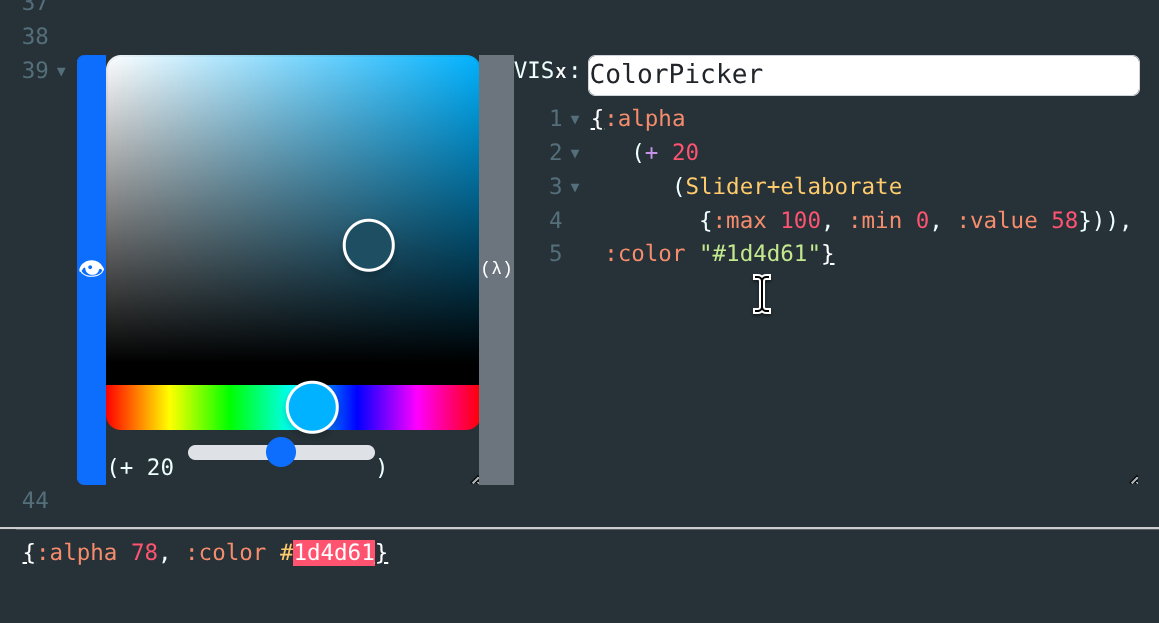}
  \caption{A color picker VIsx showing a multiple nestings of code and GUIs}
  \label{fig:color-picker}
\end{figure}

\textsc{Hybrid Clojure\-Script} programmers may occasionally wish to nest VIsx GUIs
and textual code within each other at multiple levels. While our
experiments revealed that too many nesting levels can easily lead to
incomprehensible programs, an example where this may occur in practice
is in the commonly seen "color picker" example, e.g., as shown by
Horowitz and Heer~\cite{horowitz2023liverichcomposable}. Colors in
textual code are typically represented with RGB values, hex code, or
some other unintuitive representation that does not allow seeing the
color. Thus, a VIsx that allows clicking on a desired color
visually could be more useful for
programmers. Figure~\ref{fig:color-picker} shows that such an example
can be implemented with \textsc{Hybrid Clojure\-Script}. Specifically, it depicts a color
picker where a textual code box resides inside the GUI. Inside this
code box is another GUI---a slider---that allows a programmer to use
mouse gestures instead of typing in a number. Since every VIsx
has a textual representation, they exist at the level of the program's
AST, and thus can be nested recursively to an arbitrary depth.

\section{Related work}

Our work is related to a wide body of inspirational research:
(1) languages and environments that allow programmers to run
custom programs as they edit code; (2) graphical and non-textual
programming languages; and (3) projectional and bidirectional editing.

\subsection{Edit time}

Two rather distinct pieces of work combine edit-time computation with a form of
programming. The first is found in the context of the Spoofax language workbench
project and is about general-purpose programming languages. The second,
Microsoft's mixing of textual and graphical programs in its Office
productivity suite, is domain-specific.

Spoofax~\cite{kv:spoofax} is a framework for developing programming languages.
Erdweg et al.~\cite{ekrkov:growing} recognize that, when developers grow programming
languages, they would also like to grow their IDE support. For example, a new
language feature may require a new static analysis or refactoring
transformations, and these tools should cooperate with the language's IDE. They
therefore propose a framework for creating edit-time libraries. In essence, such
libraries connect the language implementation with the IDE,
specifically the IDE tool suite. The features are
extra-linguistic, however, and thus do not support the kinds of abstraction (and meta-abstraction) enabled by VIsx extensions.

Microsoft Office plugins, called VSTO Add-ins~\cite{m:office}, allow authors to
create new types of documents and embed them into other documents. One developer
might use it to make a music type-setting editor, while another might use it to put music
notation in a PowerPoint presentation. Even though this tool set lives in the
.NET framework, it is also an extra-linguistic construct and does not allow developers to
build programming abstractions. For example, it could not be used to implement the form builder example from section~\ref{sub:meta}.

\subsection{Graphical and live languages}

Scratch~\cite{rmmrebmrssk:scratch} is a fully graphical language
system that is familiar to many novice programmers. In Scratch, users
write their programs by snapping graphical blocks together. These
blocks resemble puzzle pieces and snapping them together creates
syntactically valid programs. The language is fully graphical,
however, and thus the graphical blocks cannot be interleaved with
textual code. This can lead to scaling issues when programmers wish to
progress beyond beginner examples. Further, Scratch offers limited
(but growing) capabilities for a programmer to extend the language
with new block types~\cite{hm:bringing}.

Several programming systems have enabled a mixture of some graphical and
textual programming for decades.
Boxer~\cite{da:boxer} allows developers to embed
GUI elements within other GUI elements (``boxing''), to name such GUI elements,
and to refer to these names in program code. That is, programs consist of
graphical renderings of GUI objects and program text (inside the boxes). For
example, a Boxer programmer could create a box that contains an image of a board
game tile, name it, and refer to this name in a unit test in a surrounding box.
Boxer does \emph{not}, however,  satisfy any of the other desiderata listed in
section~\ref{sec:design-space}. In particular, it has poor support for creating
new abstractions with regard to the GUI elements, and it would not be able to show the Tsuro tile from section~\ref{sec:tsuro-intro} embedded in the code buffer.

LabVIEW~\cite{k:labview} is a commercial visual language targeted at
scientists and engineers. It is widely adopted in its target
communities. While it is possible to create robust products using
LabVIEW, it is not easily extensible, meaning that programmers cannot
easily create new types of visualizations for other domains.

Hypercard~\cite{g:complete} gives users a graphical interface to make
interactive documents. Authors have used Hypercard to create everything from
user interfaces to adventure games. While Hypercard has been used in a wide
variety of domains, it is not a general-purpose language.

Smalltalk~\cite{gr:smalltalk,iputm:lively, bcdl:deep,
kebmb:webstrates, rnaek:codestrates}, and in particular the
Sandblocks~\cite{sandblocks} implementation of Squeak described in
section~\ref{sub:compare}, has supported direct manipulation of GUI
objects, often called live programming. Rather than separating code
from objects, Smalltalk programs exist in a shared environment called
the Morphic user interface~\cite{mrw:introduction}. Programmers can
visualize GUI objects, inspect and modify their code component, and
re-connect them to the program. No conventional Smalltalk system,
however, truly accommodates general-purpose VIsx-oriented
programming as a primary mode.

GRAIL~\cite{ehs:grail,ehs:grail-project} is possibly one of the oldest
examples of graphical syntax. It allows users to create and program
with graphical flow charts. Despite the apparent limitations of this
domain, one achievement was that GRAIL was powerful enough to be
implemented using itself.  Notebooks~\cite{pg:ipython, a:observable,
w:mathematica, bcdghhlmmmov:maple} and
Webstrates~\cite{kebmb:webstrates, rnaek:codestrates} are essentially
a modern reincarnation of GRAIL, except that they use a
read-eval-print loop approach to data manipulation rather than the
GUI-based one made so attractive by the Morphic framework. These
systems do not permit domain-specific syntax extensions, however, and
the use of eval means that the ability to statically reason about
programs is reduced.

\subsection{Bidirectional and projectional editing}

Bidirectional editors attempt to present two editable views for a
program that developers can manipulate in lockstep. One example,
Sketch-n-Sketch~\cite{chsa:programmatic, hllc:deuce}, allows
programmers to create SVG-like pictures both programmatically with
text and by directly manipulating the picture. Another example is
Dreamweaver~\cite{a:dreamweaver}, which allows authors to create web
pages directly and drop down to HTML when needed. Changes made in one
view propagate back to the other, keeping them in sync. An advantage
of these systems over VIsx is that the text representation
directly produces plain code. In contrast, a \textsc{Hybrid Clojure\-Script} VIsx
text representation is serialized JSON data, from which runtime
code is generated via a macro or function.
These bidirectional systems, however, are typically
domain-specific. Further, the visual syntax in these systems are not
linguistic abstractions and these DSLs do not allow abstraction
over visual syntax, e.g., using VIsx to define new
VIsx. Our system adds VIsx to a general-purpose
programming language and preserves abstraction over new visual
extensions. Thus, we conjecture that \textsc{Hybrid Clojure\-Script} could potentially be
used as a framework to implement bidirectional DSLs like the ones
cited, as well as new ones. Dually, we plan to investigate which ideas
from bidirectional editing systems could be used to improve the
process of creating VIsx extensions in the future.

Wizards and code completion tools, such as
Graphite~\cite{oylm:active}, perform this task in one direction. A
small graphical UI can generate textual code for a
programmer. However, once finished, the programmer typically cannot
return to the graphical UI from text without re-invoking the tool and
reparsing the text.

Projectional editing aims to give programmers the ability to edit programs
visually.
In this world, developers directly edit and manipulate graphically presented abstract syntax trees (ASTs).
The system can then render the ASTs as conventional program text. The most
well-known system is MPS~\cite{psv:jetbrains, vl:supporting}. It has been used
to create large non-textual programming systems~\cite{vrsk:mbeddr}. Unlike
VIsx extensions, projectional editors must be modified in their
host editors and always demand separate edit-time and run-time modules. Such a
separation means all editors must be attached to a program project, and they cannot
be constructed locally within a file. It therefore is rather difficult to
abstract over them.

Barista~\cite{km:barista} is a framework that lets programmers mix
textual and visual program elements. The graphical extensions,
however, are tied to the Barista framework, rather than the
programs themselves. Like MPS, Barista saves the ASTs for a
program, rather than the raw text.

Larch~\cite{fkd:programs} also provides a hybrid visual-textual
programming interface. Programs written in this environment, however,
do not contain a plain-text representation. As a result, programmers
cannot edit programs made in the Larch Environment in any other
editor.

The Hazel and Livelits projects~\cite{ocmvc:livelits} also allow
programmers to embed graphical syntax into their code, and thus there
is some overlap in use cases with our system. While
section~\ref{sub:compare} presents a cursory comparison, the two
projects seem to originate from different motivations, however, which
makes a deeper comparison difficult. Livelits began with work on typed
languages and typed holes, which aim to preserve static reasoning,
i.e., type safety, in the presence of missing code. A benefit that
emerged from this capability is that the holes can be visualized to
enable more expressive visual constructs, while still preserving the
type-based guarantees. The project continues to improve and recent
work has added the ability for holes (and thus visualizations) to
appear in pattern positions~\cite{yggpmo:livepats}.

In contrast, our language mostly works with dynamic languages and thus
we do not immediately have the same benefits of type-based
reasoning. More specifically, our work originates from extensible
language work in Scheme and Racket, where static
reasoning mostly focuses on preserving lexical scope, i.e., hygiene,
even in the presence of ``towers of languages''~\cite{queinnec}, i.e., extensions
defined using other extensions. A key idea that enables creating such
towers is to separate compile time and runtime phases of
code~\cite{flatt-composable2002}. In a language with visual extensions
like ours, this requires adding a separate third phase, which is when
the rendering code for VIsx runs. Maintaining this distinction
contributes to the expressivity of visual constructs, which can
replace any language construct (not only expressions). Further, being
able to add visual syntax as extensions of a host language comes with
the adoptability benefits in this paper because programmers can
continue to program with their knowledge of the existing host
language. In contrast, most other hybrid visual capabilities, including Livelits, appear to require the use of a new language. Our system also
allows unique use cases, such as being able to define new VIsx
using existing VIsx code, which do not appear possible in other
systems.

\citet{ek:expressive} introduced an Eclipse plugin that brought
graphical elements to Java. Like VIsx, these graphical elements
have a plain-text representation, stored as Java annotations. This
implies that programmers can write code with this plugin and view it
in any plain-text editor. The plugin differs from VIsx
extensions, however, in two ways: (1) the plugins are less expressive
than elaborators; and (2) the way new types of extensions are created
limits programmers' ability to abstract over them. For example,
programmers cannot create meta-instances with this plugin.
\section{Conclusion and the way ahead}

Linear text is the most widely embraced means for writing down programs.
But, we also know that in many contexts a picture is worth a thousand
words. Developers know this, which is why ASCII diagrams accompany many
programs in comments and why type-set documentation comes with elaborate
diagrams and graphics. Developers and their support staff create these
comments and documents because they accept the ideal that code is a message
to some future programmer who will have to understand and modify it.

If we wish to combine the productivity of text-oriented programming with the
power of pictures, we must extend our textual programming languages with
graphical syntax. A fixed set of graphical syntaxes or static images do not
suffice, however. We must equip developers with the expressive power to create
interactive graphics for the problems that they are working on and integrate
these graphical pieces of program directly into the code. Concisely put, turning
geometric comments into executable code is the only way to keep them in sync
with code.

This paper describes \textsc{Hybrid Clojure\-Script}, the first language that satisfies
this criterion. Critically, it supports visual programming that is
fully compatible with its purely textual counterpart. That is,
programs written in the hybrid language can still be edited in an
otherwise purely textual editor. Further, the visual constructs are
linguistic extensions that preserve static reasoning, e.g., lexical
scope, of the program, and may be put into libraries and be used to
define other VIsx extensions.

Clearly, designing and implementing such a hybrid language is only a
beginning. The next step is to conduct research that helps developers make
decisions as to when to invest in the adaption of GUI code as a VIsx
construct. Even more importantly, the design of such constructs needs guidelines
to ensure that the VIsx construct not only deposits some textual code but that that code is sufficiently comprehensible in plain text editors such as Emacs.

\section*{Acknowledgments} This research was partially supported by the
University of Massachusetts, Boston, plus NSF grants SHF 1823244, SHF 20050550,
and SHF 2116372.

\bibliographystyle{ACM-Reference-Format}
\bibliography{paper}

\clearpage
\appendix

\section*{Appendix: Additional workflow evaluations}\label{sec:definitions}

A new insight and goal of this paper is that, in order to succeed,
hybrid languages must strive to preserve the software development
workflow of plain, text-based languages. Section~\ref{sec:evaluation}
evaluated the usability of various hybrid language systems with
respect to several major workflow operations that programmers use to
edit software. This section examines some additional, more minor
programmer actions. Specifically, the table in
figure~\ref{fig:edit-comparison-2} shows how~\citet{abf:adding}'s
design compares with the one presented in this paper and how our
system is better able to preserve programmer actions.



  \def\same{\hspace{1cm}\textasciitilde}

  \def\foo{\hbox{{\footnotesize Hyperlinking Definitions and Uses}}}
  \newdimen\stringwidth
  \setbox0=\foo
  \stringwidth=\wd0

  \def\prp#1{\hbox to \stringwidth{\footnotesize #1}}
  \def\action#1{\row{\footnotesize #1}}
  \def\row#1{\hbox to \stringwidth{#1\hfil}}

  \def\comp#1{\multicolumn{1}{c}{\footnotesize #1}}

  \def\yes#1{\hspace{1cm}\cmark$^{#1}$}
  \def\noo#1{\hspace{1cm}\xmark$^{#1}$}
  \def\yeah{\yes\relax}
  \def\nope{\noo\relax}

\begin{figure}[hbt]
  \begin{tabular}{lllll}
    \toprule
    \prp{}                   & \comp{\citet{abf:adding}}    & \comp{This paper}       \\
    {\footnotesize Programming Action} & \comp{(Racket, bespoke GUI)} & \comp{(ClojureScript, DOM)}         \\
    \midrule
    \action{Abstraction}		         & \nope	         & \yeah \\
    \action{Autocomplete} 		         & \yeah 		 & \yeah \\
    \action{Coaching} 			         & \yeah		 & \yeah \\
    \action{Code Folding} 		         & \nope 		 & \yeah \\
    \action{Comments} 			         & \same 	         & \same \\
    \action{Comparison} 			 & \nope 		 & \yeah \\
    \action{Debugging} 			         & \nope 		 & \yeah \\
    \action{Dependency Update}      		 & \yeah 		 & \yeah \\
    \action{Elimination} 	   	     	 & \nope 		 & \yeah \\
    \action\foo 	 			 & \nope 		 & \yeah \\
    \action{Merging}                 	 	 & \nope      		 & \yeah \\
    \action{Migration} 			       	 & \same 		 & \same \\
    \action{Multi-Cursor Editing} 	 	 & \same 	  	 & \same \\
    \action{Refactoring}  		       	 & \nope 		 & \yeah \\
    \action{Reflow} 			         & \yeah 		 & \yeah \\
    \action{Style} 			         & \yeah 		 & \yeah \\
    \action{Undo/Redo} 			         & \yeah 		 & \yeah \\
    \bottomrule
  \end{tabular}
  \begin{tabular}{c}
    {\same} means orthogonal to VIsx
  \end{tabular}
  \caption{Comparison of graphical syntax and minor developer workflow actions}
  \label{fig:edit-comparison-2}
\end{figure}

The minor programmer actions that are evaluated are:
\begin{itemize}

\item \emph{Abstraction} means generalizing two (or more) pieces of
 code into a single one that can then be instantiated to work in the
 original places (and more). This should even include abstraction of
 graphical syntax itself, which our VIsx feature supports.

\item \emph{Autocomplete} allows programmers to choose descriptive names and
 enter them easily; recent forms of this code action complete entire phrases of
 code. It requires semantic knowledge of the programming language. In the ideal
 case, an IDE for a hybrid language should support autocompletion of textual
 prefixes into a VIsx instance.

\item \emph{Coaching} is about the back-and-forth between programmers and
 analysis tools. A coaching tool displays the results of a (static or dynamic)
 analysis in the editor and (implicitly) requests a reaction. A simple example
 is the underlining of unbound variables; an advanced one may highlight expressions
 that force register spilling. The challenge is that adding graphical syntax
 means extending the language in a non-functional manner, and doing so comes
 with its inherent problems.

\item \emph{Code Folding} enables IDEs to hide blocks of code while editing.
 Developers like to present overviews of code with code folded. Graphical
 syntax like our VIsx must not inhibit this IDE action.

\item \emph{Comments} should be orthogonal to any graphical syntax.

\item \emph{Comparison}, often referred to as ``diff-ing'', is needed to
 comprehend small changes to existing code compared to a previous
 snapshot.  It is frequently used by both source control tools and in
 IDEs. An key advantage of VIsx having a textual equivalent is that
 code comparisons continue to work in the conventional manner.

\item \emph{Debugging} demands running a program in a step-by-step
 fashion, i.e., steps a person can move through and comprehend
 sequentially. Again, having a textual representation of VIsx
 allows debugging to continue working as it did without VIsx.

\item \emph{Dependency Update} is about updating packages and
 libraries for various reasons.  A change made to the definition of a
 VIsx is reflected in its uses automatically.

\item \emph{Elimination} is the dual of abstraction, meaning in-lining the code
 for an existing abstraction. This must continue to work, even with VIsx.

\item \emph{Hyperlinking Definitions and Uses} allows programmers to easily
 navigate between definitions and uses. To hyperlink
 pieces of code properly, an IDE must understand both the text and the
 semantics of code.  Whether this form of linking works properly depends on how
 easily an IDE can get hold of the text that corresponds to a VIsx.

\item \emph{Merging} two blocks of code is the natural extension to
 ``Comparison''. Instead of viewing a report of the difference, however, merging
 attempts to generate syntactically correct code that represents two sources
 derived from one original point. Like ``Comparison'', ``Merging'' is clearly a
 text-based action but more sophisticated. Research is needed to investigate how
 well ``Merging'' works in the presence of VIsx.

\item \emph{Migration} happens when a dependency or platform changes, breaking
 backwards compatibility. Frequently, this requires small tweaks through an
 entire codebase. Supporting extension migration suffices here.

\item \emph{Multi-Cursor Editing} allows two (or more) developers to
 concurrently edit the same program. This should be orthogonal to any
 kind of graphical syntax.

\item \emph{Refactoring} is a syntax- or even semantics-aware search-and-replace
 action. Most simple refactoring actions should work as-is even in the presence
 of VIsx. More research is needed to understand whether
 refactoring works when syntactic differences involve VIsx.

\item \emph{Reflow} automatically transforms program text in an IDE buffer to
 conform to some style standards, e.g., proper indentation. If an IDE
 accommodates VIsx, reflow continues to work.

\item \emph{Styling} changes aspects of code display, e.g., the font size or the
 color theme. Instances of VIsx may benefit from explicitly
 coordinating with style operations.

\item \emph{Undo/Redo} is straightforward for text. For VIsx, each
 extension can package multiple changes into a single undo/redo step.
\end{itemize}

\label{lastpage01}

\end{document}